%% file: main.tex
\documentclass[preprint,10pt,authoryear,5p,twocolumn]{elsarticle}

\usepackage{amssymb}
\usepackage{amsmath}
 \usepackage{amsthm}

\usepackage{physics}
\usepackage{mathtools}
\usepackage{xcolor}
\usepackage{cases}
\usepackage{enumitem} 
\usepackage{dsfont}   %
\usepackage{tikz}
\usetikzlibrary{arrows, positioning, calc}
\usepackage{circuitikz}
\usepackage{pgfplots}
\pgfplotsset{compat=newest}
\usepackage{siunitx}

\usepackage[hidelinks]{hyperref}
\hypersetup{
  pdftitle={Current trends and future directions in event-based control},
  pdfauthor={Michael Hertneck, David Meister, and Frank Allgöwer}
}

\journal{European Journal of Control}

\providecommand{\DIFdel}[1]{} %

\newcommand{\R}{\mathbb{R}}
\newcommand{\N}{\mathbb{N}}
\newcommand{\Z}{\mathbb{Z}}
\newcommand{\K}{\mathcal{K}}
\renewcommand{\L}{\mathcal{L}}
\newcommand{\C}{\mathcal{C}}
\newcommand{\D}{\mathcal{D}}

\newcommand{\E}[1]{\ensuremath{\mathbb{E}\!\left\lbrack#1\right\rbrack}}
\newcommand*\diff{\mathop{}\!\mathrm{d}}

\DeclareMathOperator\ess{ess}
\DeclareMathOperator{\e}{e}

\newtheorem{asum}{Assumption}
\newtheorem{theo}{Theorem}
\newtheorem{prop}{Proposition}

\begin{document}

\begin{frontmatter}

  \title{Current trends and future directions in event-based control} %

  \author[label1]{Michael Hertneck} %
  \author[label1]{David Meister}
  \author[label1]{Frank Allgöwer\fnref{t1}}

  \affiliation[label1]{organization={University of Stuttgart, Institute for Systems Theory and Automatic Control},%
    city={Stuttgart},
    country={Germany}}

  \begin{abstract}
    The defining characteristic of event-based control is that feedback loops are only closed when indicated by a triggering condition that takes recent information about the system into account.
    This stands in contrast to periodic control where the feedback loop is closed periodically.
    Benefits of event-based control arise when sampling comes at a cost, which occurs, e.g., for Networked Control Systems or in other setups with resource constraints.
    A rapidly growing number of publications deals with event-based control.
    Nevertheless, some fundamental questions about event-based control are still unsolved.
    In this article, we provide an overview of current research trends in event-based control.
    We focus on results that aim for a better understanding of effects that occur in feedback loops with event-based control.
    Based on this summary, we identify important open directions for future research.
  \end{abstract}


  \fntext[t1]{F.\ Allgöwer is thankful that this work was funded by the Deutsche Forschungsgemeinschaft (DFG, German Research Foundation) under Germany's Excellence Strategy -- EXC 2075.}

\end{frontmatter}

\input{intro.tex}
\input{overview.tex}
\input{past_developements.tex}

\input{frameworks.tex}

\input{sampling_behavior.tex}

\input{performance_comparison.tex}

\input{information.tex}

\input{further_trends.tex}

\input{coclusion.tex}

\bibliographystyle{elsarticle-harv}
\bibliography{literature}

\end{document}

%% file: intro.tex
\section{Introduction}
Event-based control has been proposed as an alternative to time-triggered control (TTC) for setups where sampling is associated with a cost. 
TTC closes the feedback loop based on predetermined time instants, often resulting in periodic control with constant sampling intervals.
On the contrary, when using event-based control, the feedback loop is closed only when indicated by a triggering condition, which takes recent information about the system like current state or output measurements into account.
Event-based control subsumes two main paradigms: event-triggered control (ETC) and self-triggered control (STC).
For ETC, a triggering rule based on current system information like states or outputs is monitored during the runtime of the system so that sampling instants are selected in an online fashion depending on the system state.
For STC, sampling instants are determined iteratively. 
At each sampling instant, the next instant is determined based on the current state of the system. 
Compared to ETC, STC thus avoids the continuous evaluation of a triggering rule.
In turn, STC reacts slower than ETC to unexpected effects like disturbances. 
Thus, ETC is also classified as reactive, whereas STC is classified as proactive \citep{heemels2012introduction}. 
Event-based control also includes many concepts related to ETC and STC, some of which we will discuss later.

The findings from \cite{arzen1999simple,astrom1999comparison} provided an important starting point for the research on event-based control in the past decades.
In \cite{arzen1999simple}, it is shown in simulation, that a simple threshold-based ETC triggering rule can reduce the CPU load required for control significantly with only minor degradation of the control performance compared to periodic sampling with fixed sampling period. 
In \cite{astrom1999comparison}, it is proven for first-order stochastic systems that using a threshold-based  ETC triggering rule outperforms periodic sampling with a fixed sampling period at the same average sampling rate in terms of the resulting variance. 
This is also visualized in Figures~\ref{fig_intro_TT} and \ref{fig_intro_ET}, where simulated system trajectories for a standard Wiener Process controlled using periodic sampling (Figure~\ref{fig_intro_TT}) and using a simple threshold-based ETC triggering condition (Figure~\ref{fig_intro_ET}) are plotted. 
\begin{figure}[th!]
	\centering
	\input{img/intro_example/plot_TT.tex}
	\caption{System trajectory (\ref*{fig:TT_traj}) example for standard Wiener process noise and periodic sampling interval of \SI{0.5}{\second}.
		Triggering instants are shown by vertical dashed lines (\ref*{fig:TT_events}).}
	\label{fig_intro_TT}
	\vspace*{\abovedisplayskip}
	\input{img/intro_example/plot_ET.tex}
	\caption{System trajectory (\ref*{fig:ET_traj}) example for standard Wiener process noise and event-triggered expected inter-event time of \SI{0.5}{\second}.
		Triggering instants are shown by vertical dashed lines (\ref*{fig:TT_events}).
		Event triggering utilizes a constant threshold on the system state (\ref*{fig:ET_thresh}).}
	\label{fig_intro_ET}
\end{figure}
Whilst the signal is reset to $0$ periodically in Figure~\ref{fig_intro_TT}, it is reset to $0$ whenever it reaches a fixed threshold in Figure~\ref{fig_intro_ET}. 
The threshold is chosen such that the expected inter-event time is the same for both Figures, i.e., such that the average number of sampling instants are the same. 
However, it can be seen that the variance of the signal is significantly lower for ETC compared to periodic sampling.

Both, \cite{arzen1999simple} and \cite{astrom1999comparison}, thus highlight possible advantages of event-based control and have motivated the development of systematic design approaches for event-based control mechanisms:
Event-based control has the potential to reduce the sampling rate compared to periodic control while still fulfilling closed-loop performance requirements or to reduce the control performance at the same average sampling rate.

The described potential of event-based control 
can be translated into various practical advantages:
From reducing the amount of communication in  Networked Control Systems (NCS) to its capability to save computational resources and energy in various setups \citep{astrom1999comparison,arzen1999simple,mazo2011decentralized}.
Moreover, there are also processes like satellite control with an intrinsic event-based nature that are thus well suited for event-based control \citep{astrom1999comparison,ong2024hierarchical}.

Due to its benefits, event-based control is a very active field and the number of publications in event-based control has increased almost exponentially in the recent years (documented, e.g., in numbers in the bibliometric analysis \citep{aranda2020event}).
A great boost for the research on event-based control was provided by \cite{tabuada2007event}, where a systematic design of an ETC triggering rule for a broad system class is proposed. 
After this seminal paper, many further results for event-based control with different triggering rules and a variety of methods for theoretical analysis have been developed, of which we provide an overview in a separate section. 
Despite the great research efforts in the past decades, there are still unsolved fundamental questions that need to be answered in the future.
At the same time, the rapidly increasing number of publications makes it necessary to bring structure into current research and to identify relevant research topics and differentiate them from topics that are already well-developed. 

The aim of this article is to present a focused perspective on relevant theoretical topics in event-based control in order to guide research efforts towards pressing open problems.
We provide a brief introduction in event-based control and give an overview of the existing literature.
Moreover, we summarize current research trends with a focus on theoretical results that aim for a better understanding of effects that occur in feedback loops with event-based control. 
Based on the summary, we identify important open questions and give an overview of promising directions for future research. 

The article is structured as follows.
In Section~\ref{sec_overview}, we provide a conceptual introduction to the field of event-based control and present some standard approaches from the literature.
In Section~\ref{sec_literature},  this conceptual introduction is complemented with a literature overview of the most important well-established results on event-based control.
Section~\ref{sec_framework}, contains an overview of frameworks that aim to unify classes of ETC triggering rules.
The frameworks are based on different analysis techniques and offer not only the potential to improve the understanding of existing ETC triggering rules, but also to develop new triggering rules and redesign existing ones. 
In Section~\ref{sec_sampling}, we review techniques for the analysis of the sampling behavior of systems with event-based control.
We consider approaches that allow analytical statements for relatively simple setups under certain assumptions as well as numerical approaches for more complicated setups and larger system classes.
Subsequently, in Section~\ref{sec_comparison}, we elaborate on the trade-off between performance and sampling rate faced when designing or evaluating event-based control schemes.
Section~\ref{sec_info} addresses results on the data-rate requirements for stabilizing systems over rate-limited channels when transmissions are determined using ETC. 
Section~\ref{sec_current} contains a brief summary over current research trends including results regarding new analysis techniques for ETC triggering rules, connections between ETC and other main research fields in automatic control including data-based control and machine learning, and open questions for the research on STC. 
In Section~\ref{sec_conc}, we conclude the paper by summarizing the key takeaways and outlining new domains that could profit from the event-based paradigm and the developed theory.

\subsection*{Notations}
The non-negative real numbers are denoted by  $\R_{\geq 0} $. The natural numbers are denoted by $\N$, and we define $\N_0\coloneqq\N\cup  \left\lbrace 0 \right\rbrace $. We use $(x,y) = \left[x^\top,y^\top\right]^\top$. 
A continuous function $\alpha\colon\R_{\geq 0} \rightarrow \R_{\geq 0}$ is of class $ \K$ if it is strictly increasing and $\alpha(0) = 0$. It is a class $\K_\infty$ function if it is a class $\K$ function and unbounded.
Given $t\in\R$ and a piecewise continuous function $f\colon\R\rightarrow\R^n$, we use the notation $f(t^+) \coloneqq \lim\limits_{s\rightarrow t,s>t} f(s)$.
When omitting the argument, we also abbreviate as $f^+ \coloneqq f(t^+)$.
For $x,v\in\R^n$ and locally Lipschitz $U\colon\R^n\rightarrow\R$, $U^\circ(x;v)$ denotes the generalized Clarke derivative of function $U$ at $x$ in the direction $v$, i.e., $U^\circ(x;v) \coloneqq \limsup_{y\to x, \lambda \to 0} \frac{U(y+\lambda v)-U(y)}{\lambda}$.
Furthermore, let $\E{\cdot}$ denote the expected value of a stochastic variable.

%% file: img/intro_example/plot_TT.tex
\definecolor{matlabblue}{rgb}{0.00000,0.44700,0.74100}
\definecolor{matlabred}{rgb}{0.85000,0.32500,0.09800}

\begin{tikzpicture}
	\begin{axis}[
			xlabel=Time in \SI{}{\second},
			ylabel=System state,
			width=0.95*\columnwidth,
			height=0.55*\columnwidth,
			no markers,
			xmin=0,
			xmax=8.1,
			ymin=-2,
			ymax=2
		]
		\addplot+[matlabblue] table[x=t,y=0,col sep=comma]{img/intro_example/exp_N1_T20/x_TT.csv};\label{fig:TT_traj}
		\foreach \t in {.5,1,1.5,2,2.5,3,3.5,4,4.5,5,5.5,6,6.5,7,7.5,8,8.5,9,9.5}
			{\addplot[mark=none,dashed] coordinates {(\t, -1.8) (\t, 1.8)};}\label{fig:TT_events}
		\addplot[mark=none,dotted] coordinates {(0, 0) (10, 0)};
	\end{axis}
\end{tikzpicture}

%% file: img/intro_example/plot_ET.tex
\definecolor{matlabblue}{rgb}{0.00000,0.44700,0.74100}
\definecolor{matlabred}{rgb}{0.85000,0.32500,0.09800}

\begin{tikzpicture}
	\begin{axis}[
		xlabel=Time in \SI{}{\second},
		ylabel=System state,
		width=0.95*\columnwidth,
		height=0.55*\columnwidth,
		no markers,
		xmin=0,
		xmax=8.1,
		ymin=-2,
		ymax=2,
		legend pos=south west
		]
		\addplot+[matlabblue] table[x=t,y=0,col sep=comma,forget plot]{img/intro_example/exp_N1_T20/x_ET.csv};\label{fig:ET_traj}

		\addplot[ycomb,dashed,forget plot] table[x=t_ET,y expr=1.8,col sep=comma]{img/intro_example/exp_N1_T20/t_ET.csv};
		\addplot[ycomb,dashed] table[x=t_ET,y expr=-1.8,col sep=comma]{img/intro_example/exp_N1_T20/t_ET.csv};\label{fig:ET_events}

		\addplot[mark=none,dotted,forget plot] coordinates {(0, 0) (10, 0)};
		\addplot[mark=none,dashdotted,matlabred,line width=1.0pt] coordinates {(0, -0.7071) (10, -0.7071)};\label{fig:ET_thresh}
		\addplot[mark=none,dashdotted,forget plot,matlabred,line width=1.0pt] coordinates {(0, 0.7071) (10, 0.7071)};
	\end{axis}
\end{tikzpicture}

%% file: overview.tex
\section{A brief introduction to event-based control}
\label{sec_overview}
As a basis for the paper and to give a brief introduction into event-based control, we present in this section a standard setup for ETC and STC with state feedback and present some of the most common ETC triggering rules and a simple STC approach.
\subsection{Standard setup}
\label{sec_sub_setup}
\begin{figure}
	\centering
	\input{img/fig_etc.tex}
	\caption{Block diagram for ETC with triggering mechanism that uses current state- and input information.}
	\label{fig_block_ETC}
\end{figure}
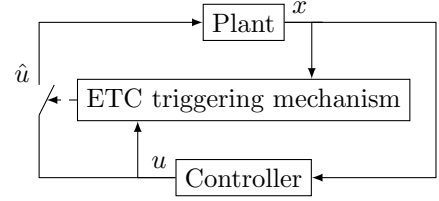

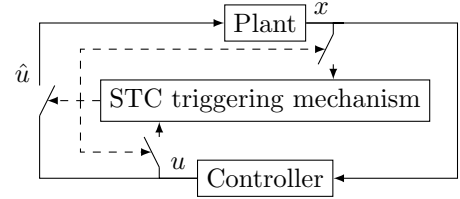
\begin{figure}
	\centering
	\input{img/fig_stc.tex}
	\caption{Block diagram for STC with triggering mechanism based on state- and input information from the last sampling instant}
	\label{fig_block_STC}
\end{figure}

We consider a plant of the form
\begin{equation}
	\label{eq_plant}
	\dot{x} = f(x,\hat{u})
\end{equation}
with $x(t)\in\R^{n_x}$ and $\hat{u}(t)\in\R^{n_u}$. 
Moreover, a state feedback controller is described by
\begin{equation}
	\label{eq_ctrl}
	u = \kappa(x)
\end{equation}
with $u(t)\in\R^{n_u}$. 
If $\hat{u}(t) = u(t)~\forall t\geq 0$, then the model \eqref{eq_plant},~\eqref{eq_ctrl} corresponds to a system with continuous state feedback. 
However, in an event-based sample-and-hold setup, the input that is applied to the plant is only updated at sampling instants. The sampling instants are determined by a triggering mechanism that is described subsequently.
We denote the sequence of sampling instants by $(t_j)_{j\in\N_0}$ with (w.l.o.g.) $t_0 = 0$ and thus obtain
\begin{equation}
	\begin{split}
		\hat{u}(t) ={}& u(t_j),~t\in\left[t_j,t_{j+1}\right)\\
	\end{split}
	\label{eq_u_etc}
\end{equation}
for all $j\in\N_0$.
For the difference between  $\hat{u}$ and $u$, we write
\begin{equation*}
	e \coloneqq \hat{u}-u
\end{equation*}
and  refer to $e$ as the \emph{sampling-induced error}.
The term sampling-induced error is chosen as it originates from using the proposed sample-and-hold setup instead of continuous feedback.
It can be concluded that $e(t_j^+) = 0$ as $\hat{u}(t_j) = u(t_j)$. 
We further define the \emph{inter-event time} as $h_j \coloneqq t_{j+1}-t_j$.
ETC and STC mainly differ in the way how sampling instants are determined.
The former continuously checks a triggering rule based, whereas the latter leverages information at sampling instants and predictions to determine the next sampling instant.
Respective block diagrams for both approaches and the presented standard setup are given in Figures~\ref{fig_block_ETC} and \ref{fig_block_STC}. 

\subsection{An overview over common ETC triggering rules}

Next, we briefly introduce common ETC triggering rules. 
These typically aim at bounding the sampling-induced error $e$ or related quantities that are suitable to describe the sampling effect.
Besides that, the triggering rules need also to be designed such, that the inter-event times $h_j$ are lower bounded for all $j$. 
This is particularly important as otherwise, Zeno behavior, i.e., an infinite number of sampling instants in a finite time interval could result from using an ETC triggering rule. 
The theoretical guarantees for the individual triggering rules, such as induced stability properties of the closed loop, are discussed later in Section~\ref{sec_framework}.

\subsubsection{Absolute threshold triggering rule}
\label{sec_sub_abs}
Absolute threshold triggering rules compare the magnitude of the sampling-induced error to a fixed constant threshold value.
Sampling instants are triggered when the constant is exceeded.  
In particular, such triggering rules have the form
\begin{equation}
	\label{eq_trigger_abs}
	t_{j+1} = \inf\left\lbrace t > t_j\mid \gamma(\abs{e(t)}) \geq \rho \right\rbrace,
\end{equation} 
for a $\K_\infty$ function $\gamma$ and a threshold parameter $\rho > 0$. 
The triggering rule thus ensures that $\gamma(\abs{e(t)}) \leq \rho$ holds for $t\in\left[t_j,t_{j+1}\right)$.

Absolute threshold triggering rules have, e.g., been used in \cite{otanez2002using,kofman2006level,miskowicz2006send}.
 
Next, an alternative triggering rule, that replaces the constant factor $\rho$ by a term that depends on the system state is discussed.

\subsubsection{Relative threshold triggering rule}
\label{sec_sub_rel}
Relative threshold triggering rules compare the magnitude of the sampling-induced error $e$ with a term that depends on the magnitude of the system state $x$.
Thus, such triggering rules have the form 
\begin{equation}
	t_{j+1} = \inf\left\lbrace t > t_j\mid \gamma(\abs{e(t)}) \geq \sigma \alpha(\abs{x(t)})\right\rbrace,
	\label{eq_trigger_rel}
\end{equation}
with $\K_\infty$ functions $\alpha$ and $\gamma$ and a threshold parameter $\sigma\in\left(0,1\right)$.
The triggering rule ensures that $\gamma(\abs{e(t)}) \leq \sigma \alpha(\abs{x(t)})$ holds for $t\in\left[t_j,t_{j+1}\right)$.
The relative threshold triggering rule was, e.g., used in \cite{tabuada2007event}.

An advantage of the relative threshold triggering rule compared to the absolute threshold triggering rule is that the origin is stabilized and not just  a set containing the origin with size that depends on the triggering parameter (stability properties will be discussed in more detail in Section~\ref{sec_framework}).
In turn, the absolute threshold triggering rule is more robust with respect to disturbances. 
In particular, arbitrary small disturbances may lead to arbitrary fast sampling for the relative threshold triggering rule.
However, this does not occur for the absolute threshold triggering rule \citep{borgers2014event}.  
Both triggering rules can also be combined to a mixed threshold triggering rule to achieve a trade-off between the properties of both strategies \citep{donkers2012output,borgers2014event}. 
Such a combination is also referred to as \emph{space regularized relative threshold ETC} and can have decent robustness properties with respect to measurement noise \citep{scheres2024robustifying}. 

Next, we present a dynamic triggering rule that extends the relative threshold triggering rule by a dynamic variable.

\subsubsection{Dynamic triggering rule}
Dynamic triggering rules were proposed for the first time in \cite{girard2015dynamic,mousavi2015integral}. The idea is to add internal dynamics to the triggering rule. 
To that end, a dynamic variable $\eta(t)\in\R$ with dynamics
\begin{equation}
	\label{eq_def_dyn}
	\dot{\eta} = -\beta(\eta) + \sigma \alpha(\abs{x}) - \gamma(\abs{e}), ~\eta(0) = 0
\end{equation}
with $\beta\in\K_\infty$ and $\sigma\in\left(0,1\right)$ is introduced. 
The dynamic triggering rule is given by
\begin{equation}
	t_{j+1} = \inf\left\lbrace t > t_j\mid \eta(t) \leq 0 \right\rbrace
	\label{eq_trigger_dyn}
\end{equation} 
and ensures that $\eta(t) \geq 0$ holds for $t\in\left[t_j,t_{j+1}\right)$.

The dynamic triggering rule can be interpreted as an averaged version of the relative threshold triggering rule. 
In particular, it allows choosing sampling instants such that the respective relative threshold triggering rule is violated for some time as long as it is still satisfied on average. 
Moreover, the dynamic triggering rule can easily be modified for setups where additional effects like delays or packet loss occur by modifying the update of the dynamic variable in a suitable way.
In turn, memory and suitable computational capabilities for the update of the dynamic variable are required to implement the dynamic triggering rule.

Another alternative is to formulate triggering rules that directly use the Lyapunov function as we present next.

\subsubsection{Triggering based on a Lyapunov function}
\label{sec_overview_lyap}
In \cite{wang2008event}, a triggering rule is proposed that enforces a certain decrease of a Lyapunov function. The triggering rule has the form
\begin{equation}
	\label{eq_trigger_lyap}
	t_{j+1} = \inf\left\lbrace t > t_j\mid V(x(t)) \geq (1- \sigma(t-t_j)) V(x(t_j)) \right\rbrace,
\end{equation}
where $V$ is a Lyapunov function that needs to satisfy certain properties and  $\sigma \in \left(0,1\right)$ is a design parameter.
A new sampling instant is thus triggered if the Lyapunov function $V$ exceeds a tunable threshold.
This is meant to recursively yield an upper bound on $V$.  
Note that the time dependent part of the triggering rule \eqref{eq_trigger_lyap} can also be modeled in terms of a dynamic variable. 
Thus, this triggering rule can also be interpreted as dynamic triggering rule. 

An advantage of the triggering rule based on a Lyapunov function is, that it allows to directly design the system performance in terms of the convergence rate of the Lyapunov function $V$. 
However, when disturbances occur, it may be infeasible or not desirable to enforce a specific convergence rate of the Lyapunov function $V$.

After having presented some basic concepts for ETC, we will next look at an introductory STC approach. 

\subsection{Self-triggered control}
\label{sec_stc}
The concept of STC, initially proposed in \cite{velasco2003self}, is to determine at each sampling instant, when the next sampling instant should take place.
This is typically achieved using predictions for the future behavior of the state of the plant. 
Based on these predictions, a lower bound for the time when an ETC triggering rule would trigger the next sampling instant is determined, so that stability guarantees from ETC are preserved.

To illustrate this, let us consider linear systems with $f(x,\hat{u}) = Ax+Bu$ and $\kappa(x) = K x$ for matrices $A, B$ and $K$ of suitable dimensions.
In this setup, the exact solution for $x$ can be determined and used to lower bound the time, when an ETC triggering rule would trigger the next sampling instant.
In particular, 
we obtain for any specific $\delta \in\left[0, t_{j+1}-t_j\right]$ that
\begin{equation*}
	x(t_j+\delta) = G(\delta) x(t_j), 
\end{equation*}
where $G(\delta) \coloneqq e^{A \delta}  + \int_{0}^{\delta} e^{A(\delta-\tau)} BK\mathrm{d}\tau$ holds.
Using this and $e(t) = u(t_j)-u(t) = K(x(t_j)-x(t))$, \eqref{eq_trigger_rel} can be expressed as 
\begin{equation*}
	\begin{split}
		t_{j+1} ={} t_j + \inf\lbrace &\delta > 0\mid \\ &\gamma(\abs{K(I-G(\delta)) x(t_j)}) \geq \sigma \alpha(\abs{G(\delta) x(t_j)})\rbrace
	\end{split}	 
\end{equation*}
Thus, the next inter-event time $h_j$ when using triggering rule \eqref{eq_trigger_rel} is the smallest positive value of $\delta$, for which 
\begin{equation}
	\label{eq_trigger_delta}
	\gamma(\abs{K(I-G(\delta)) x(t_j)}) \geq \sigma \alpha(\abs{G(\delta) x(t_j)}) 
\end{equation}
holds.
Due to the contained matrix exponentials, it is however not feasible to compute the respective value explicitly based on \eqref{eq_trigger_delta}.
Instead, \eqref{eq_trigger_delta} can be checked for any given value of $\delta$.
This can be done for different candidate values.
The largest candidate value for which \eqref{eq_trigger_delta} does not hold is then selected as next sampling instant.
This results in a trade-off between accuracy and computational effort.  
This procedure yields an STC mechanism that always provides a lower bound for the next sampling instant that would result from the relative threshold triggering condition \eqref{eq_trigger_rel}. 

Note that the above triggering mechanism is inspired by the one from \cite{mazo2010iss}, where a Lyapunov function based triggering rule is evaluated exactly for different candidates for the next sampling instant. 
This yields a similar trade-off between accuracy and computational effort as the above mechanism.

Of course, different ways to lower bound the time, when an ETC triggering rule would trigger are possible as well. 
For example, Lipschitz continuity can be used to bound the evolution of the system state.
The main advantage of STC compared to ETC is, that no triggering rule needs to be monitored continuously for STC.
In turn, STC reacts slower to unexpected effects like disturbances (or needs to act more conservatively and thus trigger more often than ETC), as the control is only changed at the next sampling instant. 
Moreover, the used prediction may be conservative or computationally expensive (particularly when nonlinear systems are considered)

While ETC and STC, which we have briefly discussed in this section, are among the most common event-based control strategies, there are also many other event-based control strategies.
Some of them will be mentioned in Section~\ref{sec_literature}, e.g., periodic event-triggered control (PETC).
After this short introduction to event-based control we will give a broader overview of results for event-based control in the next section.

%% file: img/fig_etc.tex
\begin{tikzpicture}[node distance=.5cm, auto, , >=latex]
	\node (plant) [draw, rectangle] {Plant};
	\node (trigger) [draw, rectangle, below=of plant] {ETC triggering mechanism};
	\node (controller) [draw, rectangle, below=of trigger] {Controller};
	
	\coordinate[left=of trigger] (switch);
	\coordinate[above= 2mm of switch] (switchup);
	\coordinate[below= 2mm of switch] (switchdown);
	\coordinate[right= 2mm of switchup] (switchupright);
	\coordinate[right= 1mm of switch] (switchtrigger);
	\coordinate[above =0mm of trigger] (triggerinpa);
	\coordinate[right =9mm of triggerinpa] (triggerinpb);
	\coordinate[below =0mm of trigger] (triggerinpc);
	\coordinate[left =14mm of triggerinpc] (triggerinpd);
	
	\draw[->] (plant.east) -- node[pos = 0.1, above] {$x$}  ++(2,0) |-  (controller.east);
	\draw[-] (switchupright) -- (switchdown);
	\draw[-] (controller.west) -| node[pos = 0.06, above] {$u$} (switchdown);
	\draw[->] (switchup) |- node[pos = 0.1, left] {$\hat{u}$} (plant.west);
	\draw[->,dashed] (trigger.west) -- (switchtrigger);
	\draw[->] (plant.east) -- ++(0.5,0) -| (triggerinpb);
	\draw[->] (controller.west) -- ++(-0.5,0) -| (triggerinpd);
\end{tikzpicture}

%% file: img/fig_stc.tex
\begin{tikzpicture}[node distance=.5cm, auto, >=latex]
	
	\node (plant) [draw, rectangle] {Plant};
	\node (trigger) [draw, rectangle, below=of plant] {STC triggering mechanism};
	\node (controller) [draw, rectangle, below=of trigger] {Controller};
	
	\coordinate[left= .8cm of trigger] (switch);
	\coordinate[above= 2mm of switch] (switchup);
	\coordinate[below= 2mm of switch] (switchdown);
	\coordinate[right= 2mm of switchup] (switchupright);
	\coordinate[right= 1mm of switch] (switchtrigger);
	\coordinate[above =0mm of trigger] (triggerinpa);
	\coordinate[right =9mm of triggerinpa] (triggerinpb);
	\coordinate[below =0mm of trigger] (triggerinpc);
	\coordinate[left =14mm of triggerinpc] (triggerinpd);
	
	\coordinate[above =6mm of triggerinpb] (switchpup);
	\coordinate[above =2mm of triggerinpb] (switchpdown);
	\coordinate[left= 2mm of switchpdown] (switchpdownleft);
	\coordinate[above =4mm of triggerinpb] (switchp);
	\coordinate[left= 1mm of switchp] (switchptrigger);
	
	\coordinate[below =6mm of triggerinpd] (switchcup);
	\coordinate[below =2mm of triggerinpd] (switchcdown);
	\coordinate[left= 2mm of switchcdown] (switchcdownleft);
	\coordinate[below =4mm of triggerinpd] (switchc);
	\coordinate[left= 1mm of switchc] (switchctrigger);
	
	\coordinate[left= 2.6mm of trigger.west] (triggersource);
	
	\draw[-] (switchpdownleft) -- (switchpup);
	\draw[-] (switchcdownleft) -- (switchcup);
	
	\draw[->] (plant.east) -- node[pos = 0.1, above] {$x$}  ++(2,0) |-  (controller.east);
	\draw[-] (switchupright) -- (switchdown);
	\draw[-] (controller.west) -| node[pos = 0.06, above] {$u$} (switchdown);
	\draw[->] (switchup) |- node[pos = 0.1, left] {$\hat{u}$} (plant.west);
	\draw[->,dashed] (trigger.west) -- (switchtrigger);
	\draw[-] (plant.east) -- ++(0.5,0) -| (switchpup);
	\draw[->] (switchpdown) -- (triggerinpb);
	\draw[->] (switchcdown) -- (triggerinpd);
	\draw[-] (controller.west) -- ++(-0.5,0) -| (switchcup);
	\draw[->,dashed] (triggersource) |- (switchptrigger);
	\draw[->,dashed] (triggersource) |- (switchctrigger);
\end{tikzpicture}

%% file: past_developements.tex
\section{Overview over the literature on event-based control}
\label{sec_literature}
Over the past decades, the theory for event-based control has been developed in many directions. 
This includes various aspects such as deriving different triggering rules, different ways of analyzing known triggering rules, and the consideration of effects that occur in practical applications of event-based control like in NCS. 
In this section, our aim is to briefly summarize some of the most important concepts and problem setups regarding event-based control and to take a broadened view on the field.
Whilst the section is intended to give an overview of the concepts in the literature,
it is by no means complete in the sense that it covers the entire literature on event-based control. 
We structure this overview in subsections.
However, this separation is to be understood as a red thread for this section and not as a strict categorization for the mentioned works.
Moreover, we focus on the most recent research trends in the subsequent sections and therefore do not include them in this literature review.

\subsection{ETC analysis and design}

Various techniques to design and analyze stability properties of ETC triggering rules have been proposed in the literature. 
In \cite{heemels2008anaylsis}, a set-based triggering rule is presented. 
Sampling occurs when the system state leaves a specified set.
The triggering rule from  \cite{heemels2008anaylsis} is thus related to the absolute threshold triggering rule and has similar properties. 
An ETC approach that bounds the deviation of the trajectories of the closed-loop system relative to simulated trajectories of the closed-loop system with continuous feedback with an absolute threshold triggering rule is proposed in \cite{lunze2010state}.
This has the benefit that the maximum deviation from the trajectories of the closed-loop system with continuous feedback can be tuned as a performance measure.
However, this can also be a restriction, as deviating trajectories may also be suitable to achieve control goals. 
An ETC triggering rule with an exponentially decreasing threshold is introduced in \cite{seyboth2013event}.
This triggering rule can straightforwardly be implemented in distributed setups based on locally available information.
Time-delay methods are utilized for the design of ETC triggering rules in \cite{peng2013novel,peng2013event}.
Using these methods allows an efficient analysis and the co-design of triggering rule and feedback law.
The co-design of triggering rule and feedback law is also studied in  \cite{abdelrahim2018co} using hybrid system techniques.
Such co-designs aims to guarantee a certain performance level and to reduce at the same time the sampling frequency.
However, certifying such strict improvements is typically challenging.

A passivity-based approach for the analysis of the relative threshold triggering rule is presented in \cite{yu2013event}. 
For this approach, passivity properties of plant and controller are leveraged, which allows to efficiently analyze many additional effects like disturbances, quantization or delays. 
In \cite{liu2015small}, an input-to-state stability (ISS) condition in the form of an $\L_\infty$-norm bound is considered for the analysis of relative and absolute threshold triggering rules. 
This provides an alternative small-gain inspired ISS perspective compared to the Lyapunov function based ISS arguments from \cite{tabuada2007event}.
It is however not clear if one or the other perspective is superior to design ETC triggering rules.
Note that it is in general not trivial to extend ETC triggering rules to output feedback, as guaranteed lower bounds on the inter-event times are often not preserved under output feedback. 
This has been studied, e.g., in \cite{lehmann2011event} using a Luenberger observer with continuous feedback, in \cite{donkers2012output}  for a mixed threshold triggering rule using an impulsive system model for linear system, and in \cite{Tallapragada2012,abdelrahim2016stabilization,dolk2017output} using a time regularized triggering rule, that only triggers sampling after a minimum inter-event time has lapsed. 
The triggering rules in \cite{abdelrahim2016stabilization,dolk2017output} are based on a dynamic clock variable, a technique adapted from the analysis of minimum sampling intervals.

An approach to derive static and dynamic ETC triggering rules based on Riccati equations is presented in \cite{borgers2018riccati}. 
ETC with a triggering rule based on reachable sets is studied in \cite{brunner2019event}.
This triggering rule ensures that all reachable states are contained in some desired set. 
In \cite{ong2024performance}, a triggering rule that uses a performance barrier function is proposed. 
Sampling instants are triggered such that the performance of the system in terms of the convergence rate of a Lyapunov function satisfies a tunable bound.  
This is related to the triggering rule based on a Lyapunov function from Section~\ref{sec_overview_lyap} but offers additional flexibility in the design of the performance bound. 
In \cite{proskurnikov2020lyapunov}, it is shown that if a control Lyapunov function with some predefined convergence rate is given, then the same rate can be achieved by ETC and STC. 
The concept of periodic event-triggered control (PETC) is proposed in \cite{heemels2013periodic}. 
Instead of evaluating the triggering rule continuously, it is only evaluated at fixed sampling instants, leading to a trade-off between ETC and TTC. 
Potential benefits of PETC include the simpler implementation on digital hardware as well as a guarantee of a minimum inter-event time by design.
PETC for nonlinear systems is studied, e.g.,  in \cite{borgers2018periodic,wang2020periodic}.

\subsection{ETC effects and setup variants}

As we have already commented in Section~\ref{sec_sub_rel}, ETC triggering rules may be sensitive to external disturbances leading to unwanted phenomena like arbitrarily fast sampling.
Thus, it is important to explicitly analyze the robustness of ETC triggering rules with respect to disturbances. 
Concepts to robustify ETC in the presence of disturbances include time regularization \citep{forni2014event,abdelrahim2016stabilization,abdelrahim2017robust,borgers2018riccati}, using design tools for hybrid systems \citep{chai2020analysis} and event-holding control \citep{wang2019state}.
Further approaches with robustness to disturbances are presented in \cite{yu2013event,liu2015small,rahnama2018passivity} using disturbance bounds. 
The $\L_p$ stability of ETC triggering rules is studied in \cite{dolk2017output}.

In many papers, effects that occur in practical applications of ETC are analyzed.
Delays can, e.g.,  occur in NCS or be caused by the time it takes to compute the control law. 
While computation delays were already considered in \cite{tabuada2007event,wang2008event}, further analyses are carried out, e.g., in \cite{lehmann2012event,yu2013event,garcia2013model,dolk2017output,borgers2018riccati,rahnama2018passivity}.
ETC triggering rules with robustness to packet loss are studied, e.g., in \cite{wang2011event,lehmann2012event,dolk2017event,rahnama2018passivity}.
A number of ETC schemes like \cite{garcia2013model,yu2013event,tanwani2016observer,liu2018event,rahnama2018passivity} also explicitly consider quantization effects, that mainly occur in NCS due to the limited data rate occurring there. 
Measurement noise is considered in \cite{abdelrahim2017robust,scheres2024robustifying}.

In many setups, sensors are not collocated with each other.
For such setups, distributed and decentralized ETC approaches with local triggering rules can be used at the different sensors relying only on locally available information.
Such triggering rules are, for example, presented in \cite{mazo2011decentralized,wang2011event,tallapragada2014decentralized,dolk2017output}.
Event-triggered control for multi-agent consensus is, e.g., studied in \cite{dimarogonas2012distributed,seyboth2013event}. 
A survey on event-triggered coordination of networked systems with a particular focus on multi-agent consensus is given in \cite{Nowzari2019}.
Further, event-triggered state estimation, where a remote observer obtains state information only sporadically at sampling times that are determined by a triggering rule is, e.g., discussed in \cite{li2010event,trimpe2014event}.

In \cite{lunze2010state,garcia2013model}, the concept of model-based ETC is proposed. 
The behavior of the plant is simulated at the actuator, which has the potential of greatly reducing the number of sampling instants in NCS setups. 
Event-triggered control for trajectory tracking is studied in \cite{tallapragada2013event}.
Using ETC to save resources in the context of model predictive control (MPC) and to leverage benefits of MPC is for example studied in \cite{li2014event,hashimoto2017event,bunner2017robust}.
The concept of rollout ETC, where triggering times and control inputs are jointly optimized in a receding horizon fashion is introduced in \cite{antunes2014rollout}.

Instead of a control goal, estimation problems can also be addressed in an event-triggered fashion.
Examples for works on event-based estimation and observer design are \cite{Sijs2009,Trimpe2012,Wu2013,Mohammadi2017,scheres2021event,petri2024decentralized}.

\subsection{STC concepts}

There are also various approaches for STC in the literature, that are based on different ETC triggering rules and that use different ways to determine a lower bound on the triggering times for the ETC triggering rule. 
STC mechanisms with more sophisticated predictions based on Lipschitz continuity properties compared to the one presented in Section~\ref{sec_stc} are, e.g., given in \cite{tiberi2013simple,theodosis2018self}.
Both mechanisms determine a lower bound on the time when an absolute threshold ETC triggering rule would trigger.
Relative threshold triggering rules in combination with Lipschitz continuity properties are, e.g., considered in \cite{wang2009self,liu2015small}.
An estimate based on a Taylor approximation is used in \cite{dibenedetto2013digital}. 

An STC strategy for linear systems based on the expected evolution for a quadratic cost function is presented in \cite{gommans2014self}.  
Homogeneity is leveraged in  \cite{anta2010sample} to approximate a relative threshold triggering rule. 
This is further developed for general nonlinear systems using approximations for isochronous manifolds in \cite{anta2012exploiting} and based on finite-state abstractions in \cite{delimpaltadakis2021isochronous,delimpaltadakis2021region}.
STC based on reachable sets for linear systems is considered in \cite{brunner2019event}. 
A hybrid Lyapunov function approach is used in \cite{hertneck2024robust}.

As pointed out in Section~\ref{sec_stc}, on the one hand, disturbance attenuation properties of STC are typically worse than for ETC if the presence of disturbances is not considered in the triggering rule, as for example in \cite{mazo2010iss}.
This can be addressed by incorporating disturbance properties in the triggering rule, see, e.g., \cite{brunner2019event,Gleizer2020,Pouthier2025}.
However, considering the presence of disturbances in the prediction used in the triggering rule naturally leads to more conservative triggering of STC and thus earlier triggering compared to ETC.

A concept related to self-triggered control is minimum attention control, where an optimization problem is formulated to find inputs that maximize the inter-event time, studied in \cite{brockett1997minimum,donkers2012minimum}. 

\subsection{Summary}

Finally, introductions and overviews of event-based control can be found, e.g., in \cite{lemmon2010event,heemels2012introduction,miskowicz2015event,heemels2021event,Postoyan2025}.

Examples for practical/real-world applications of event-based control (or estimation) can be found in robotics \citep{Trimpe2011,Postoyan2015,Cuenca2019}, greenhouse automation \citep{Pawlowski2016}, vehicle platooning \citep{Dolk2017}, event-based energy metering \citep{Simonov2017}, and more.
Further application domains can also be found in Sections~\ref{sec_current} and \ref{sec_conc}, where we discuss trends in event-based control and promising research fields to be explored.

As can be seen, many concepts and analysis techniques are available for event-based control. 
Which one to use is a matter of the specific problem setup and the goal to be achieved. 
However, there are still open issues and limitations present in the above approaches.

Following the overview of well established results in this section, we will present in the subsequent sections more recent results that are related to current research problems for event-based control that allow a more unified view on the problem.

%% file: frameworks.tex
\section{Comprehensive analysis frameworks for ETC}
\label{sec_framework}
In Sections~\ref{sec_overview} and \ref{sec_literature}, we have discussed different common ETC triggering rules. 
To analyze the stability properties of ETC triggering rules, different approaches exist. 
For most triggering rules, initially specific analysis techniques tailored for a specific triggering rule were developed.
Moreover, many other analysis techniques for specific triggering rules have been developed, some of which we have briefly summarized in Section~\ref{sec_literature}.

Besides finding alternative approaches to derive or analyze specific triggering rules, it is also desirable to unify proof techniques for preferably broad classes of triggering rules to be able to make more general statements.
In particular, it is desirable to aim at frameworks that allow to embed different triggering rules for the analysis of the properties of closed-loop systems with ETC.
Such frameworks cannot only be used for the stability analysis, but in addition also facilitate understanding the underlying mechanisms and relationships between mechanisms as well as advantages and disadvantages of triggering rules.
Moreover, they may help to develop new improved triggering rules that leverage flexibility in the design. 
In this section, we give an overview of unifying frameworks to derive ETC triggering rules and explain their benefits. 
We further discuss the possibility for the development of even more general frameworks. 

The frameworks in this section are emulation-based.
This means, first the controller $\kappa$ is synthesized assuming continuous feedback such that it ensures suitable robustness properties with respect to the sampling-induced error $e$. 
Then these robustness properties are leveraged for the analysis and design of triggering rules.

\subsection{A framework based on an ISS Lyapunov function}
The first framework that we present uses an ISS condition in the form of an ISS Lyapunov function as robustness property in the emulation process.
Such an approach is used in many papers where specific triggering rules are studied. 
For example, it was used in \cite{tabuada2007event} to study the relative threshold triggering rule. 
In particular, the framework is build on the assumption, that the following condition holds.
\begin{asum}
	\label{asum_iss}
	There is an ISS Lyapunov function $V:\R^{n_x}\rightarrow \R_{\geq 0}$ such that 
	\begin{align}
		\underline{\alpha}(\abs{x})\leq V(x) \leq{}& \overline{\alpha}(\abs{x})\\
		\frac{\partial V(x)}{\partial x}  f(x,\kappa(x)+e) \leq{}& -\alpha(\abs{x}) + \gamma(\abs{e}) \label{eq_lyap_dec}
	\end{align}
	holds for suitable $\underline{\alpha}, \overline{\alpha}, \alpha, \gamma \in\K_\infty$.
\end{asum} 
Assumption~\ref{asum_iss} essentially requires that the closed-loop system has some robustness with respect to the sampling-induced error $e$.
The robustness is quantified in \eqref{eq_lyap_dec} through the functions $\alpha$ and $\gamma$.
Leveraging this robustness, the stability for specific triggering rules and the existence of lower bounds on the inter-event times are studied by calculating the time derivative of the ISS Lyapunov function $V$. 

\subsubsection*{Application of the framework to specific triggering rules}
Recall that the absolute threshold triggering rule \eqref{eq_trigger_abs} ensures that  
$\gamma(\abs{e(t)}) \leq \rho$ holds for $t\in\left[t_j,t_{j+1}\right)$. 
If in addition, the inter-event times $h_j$ do not vanish for any $j$, then it follows from  \eqref{eq_lyap_dec} that
\begin{align*}
	\frac{\partial V(x)}{\partial x}  f(x,\kappa(x)+e) \leq{}& -\alpha(\abs{x}) + \gamma(\abs{e}) \\
	\leq{}&-\alpha(\abs{x}) + \rho 
\end{align*}
holds, i.e., the Lyapunov function $V$ is decreasing outside the set $\left\lbrace x\in\R^{n_x}\mid\abs{x} \geq \alpha^{-1}(\rho) \right\rbrace$. 
The fact that the inter-event times $h_j$ do not vanish for any $j$ can be concluded for the absolute threshold triggering rule under additional Lipschitz continuity assumptions. 
Thus, the set $\left\lbrace x\in\R^{n_x}\mid V(x) \leq \underline{\alpha}\left(\alpha^{-1}(\rho)\right) \right\rbrace$ is asymptotically stable for system \eqref{eq_plant} with input generated according to \eqref{eq_ctrl}, \eqref{eq_u_etc} and triggering rule \eqref{eq_trigger_abs}. 
The constant $\rho$ can be used to determine the size of this set.

The relative threshold triggering rule \eqref{eq_trigger_rel} ensures  that $\gamma(\abs{e(t)}) \leq \sigma \alpha(\abs{x(t)})$ holds.
Plugging this condition into \eqref{eq_lyap_dec}, we observe that
\begin{align*}
	\frac{\partial V(x)}{\partial x}  f(x,\kappa(x)+e)
	\leq{}& -\alpha(\abs{x}) + \gamma(\abs{e}) \\
	\leq{}&-(1-\sigma)\alpha(\abs{x}) < 0,
\end{align*}
holds if in addition, the inter-event times $h_j$ do not vanish for any $j$. %
Then, the Lyapunov function $V$ decreases along solutions of the system.
Lower bounds on the inter-event times are provided in \cite{tabuada2007event}  under additional Lipschitz continuity assumptions.
We can thus conclude that the origin $x = 0$ is asymptotically stable for system \eqref{eq_plant} with input generated according to \eqref{eq_ctrl}, \eqref{eq_u_etc} and triggering rule \eqref{eq_trigger_abs}. 

For the dynamic triggering rule \eqref{eq_trigger_dyn}, $\eta(t) \geq 0$ holds for $t\in\left[t_j,t_{j+1}\right)$. Considering the Lyapunov function $W(x,\eta) = V(x) + \eta$, we obtain for $t\in\left[t_j,t_{j+1}\right)$
\begin{align*}
	&\frac{\partial W(x,\eta)}{\partial x} f(x,\kappa(x)+e) + \frac{\partial
		W(x,\eta)}{\partial \eta} \dot{\eta}
	\\
	\leq{}& -\alpha(\abs{x(t)}) + \gamma(\abs{e(t)})  -\beta(\eta(t)) + \sigma \alpha(\abs{x(t)}) - \gamma(\abs{e(t)})\\
	\leq{}& (1-\sigma) \alpha(\abs{x(t)}) -\beta(\eta(t))<0.
\end{align*}
Moreover, a bound on the inter-event times follows directly from the bound for the relative threshold triggering rule. 
Thus, the Lyapunov function $W$ decreases along solutions of the system. 
Hence, the origin $x = 0,\eta = 0$ is asymptotically stable for system \eqref{eq_plant} with input generated according to \eqref{eq_ctrl}, \eqref{eq_u_etc}, triggering rule \eqref{eq_trigger_dyn} and dynamic variable \eqref{eq_def_dyn}.
In contrast to the relative threshold triggering rule, the dynamic triggering rule does not require the Lyapunov function $V$ to decrease at all times. 
Instead, only on average a decrease is required, thus relaxing the condition. 

For the triggering rule based on a Lyapunov function \eqref{eq_trigger_lyap}, stability follows directly from the upper bound on the Lyapunov function that results from recursive application of the triggering rule.
Hence, the main issue is to guarantee that inter-event times do not vanish. 
This can be guaranteed if the system is exponentially ISS with respect to the sampling-induced error, i.e., if Assumption~\ref{asum_iss} holds with  $\alpha(\abs{x}) > \sigma V(x)$ and quadratic functions $\underline{\alpha}$ and $\overline{\alpha}$. 

Whilst the above described framework is based on relatively well-known techniques through the use of an ISS Lyapunov function, a separate proof of stability must still be provided for each triggering rule. 
In the next framework we present, this is generalized by using a hybrid systems framework.

\subsection{A unifying Lyapunov framework  based on hybrid systems}
A unifying framework based on hybrid systems and Lyapunov analysis is proposed in \cite{postoyan2015framework}.  
Therein, the hybrid systems formalism from \cite{goebel2012hybrid} is used to model the closed-loop system with ETC.
Thus, a hybrid system of the form
\begin{equation}
	\label{eq_hyb}
	\begin{cases}
		\dot{\xi} = F(\xi), & \xi\in\C,\\
		\xi^+ = G(\xi), & \xi\in \D,
	\end{cases}
\end{equation}
is considered, where $\xi$ is the state of the system, $F$ describes flow dynamics, $G$ describes jump dynamics, and $\C$ and $\D$ are the flow and jump sets.
Hybrid systems of the form \eqref{eq_hyb} essentially behave in such a way that they satisfy the flow dynamics as long as the state $\xi$ is in the flow set, and jump according to the jump dynamics when $\xi$ is in the jump set. 
If $\xi$ is in the intersection of both sets, the system can either flow or jump. 
More formal explanations and definitions on hybrid systems can be found in \cite{goebel2012hybrid}.
To model the system with ETC triggering rules as hybrid system of the form \eqref{eq_hyb}, the flow dynamics are used to describe the behavior of the system between sampling instants, the jump dynamics are used to describe the update at sampling instants.
Moreover, the flow and jump set represent the triggering rule, i.e., $\C$ and $\D$ are directly tied to the occurrence of events.
In particular, the choice
$\xi=(x,e,\eta)$
 with
\begin{equation*}
	F(\xi) = \begin{bmatrix}
		f(x,\kappa(x)+e)                                          \\
		- \frac{\partial \kappa(x)}{\partial x} f(x,\kappa(x)+e)  \\
		g(x,e,\eta)
	\end{bmatrix}, \quad%
	G(\xi) = \begin{bmatrix}
		x \\
		0 \\
		\ell(x,e,\eta)
	\end{bmatrix},
\end{equation*}
can be made.
Here, $\eta(t) \in\R^{n_\eta}$ contains the dynamic variables that occur for dynamic triggering rules with dynamics described by $g$ and $\ell$. 

For hybrid systems of the form \eqref{eq_hyb}, a Lyapunov condition is established in \cite[Theorem~1]{postoyan2015framework}.  The condition essentially requires that a Lyapunov function $R:\C\cup\D\cup G(\D) \rightarrow \R_{\geq 0}$ satisfies the following conditions:
\begin{enumerate}[label=(\roman*)]
	\item The function $R$ is lower and upper bounded by $\K_\infty$ functions.
	\item The generalized directional derivative of $R$ decreases along solutions on the flow set (except for the set to be stabilized), i.e., $ R^\circ(\xi;F(\xi)) < 0$ for almost all $\xi\in\C$ if the goal is to guarantee asymptotic stability of the origin. 
	\item At jumps, $R$ does not increase, i.e., $R(G(\xi)) \leq R(\xi)~ \forall \xi \in \D$ .
	\item There is a non-vanishing time between jumps of the hybrid system \eqref{eq_hyb}.
\end{enumerate}
If the conditions hold, stability of the origin can be concluded, cf.\ \cite[Theorem~1]{postoyan2015framework}.
The conditions can also be used to show stability of a set containing the origin, which is, for example, used when analyzing the absolute threshold triggering rule. 
A benefit of the framework based on hybrid systems is, that the Lyapunov function does not need to be continuous everywhere. For example, it may be discontinuous when the system jumps. Technical details are omitted here but can be found in \cite[Theorem~1]{postoyan2015framework}. 

\subsubsection*{Application of the framework to specific triggering rules}
Various different ETC triggering rules from the literature are studied in \cite{postoyan2015framework} and choices for $\C$ and $\D$ for different triggering rules are presented, so that stability can be concluded.

For the relative threshold strategy from Section~\ref{sec_sub_rel}, the choice
\begin{equation}
	\C=\left\lbrace \xi\mid \gamma(\abs{e}) \leq \sigma\alpha(\abs{x}) \right\rbrace,~\D = \left\lbrace \xi\mid \gamma(\abs{e}) \geq \sigma\alpha(\abs{x}) \right\rbrace,
	\label{eq_flow_jump_rel}
\end{equation}
can be used.
Then, the Lyapunov function $R(\xi) = \max\left\lbrace V(x), \gamma(\abs{e})\right\rbrace$ can be used to conclude stability. 

The absolute threshold strategy from Section~\ref{sec_sub_abs} can be analyzed using  
\begin{equation}
	\C=\left\lbrace \xi\mid \gamma(\abs{e}) \leq \rho \right\rbrace,~\D = \left\lbrace \xi\mid \gamma(\abs{e}) \geq \rho \right\rbrace.
	\label{eq_flow_jump_abs}
\end{equation}
The Lyapunov function $$R(\xi) = \max\left\lbrace V(x) - \alpha^{-1}(2\rho),0 \right\rbrace + \max\left\lbrace \gamma(\abs{e}) - \rho, 0\right\rbrace$$ can be used in this case to show that the set $$\left\lbrace \xi\mid V(x) \leq \underline{\alpha}\left(\alpha^{-1}(\rho)\right)\wedge \gamma(\abs{e}) \leq \rho\right\rbrace$$ is stabilized. 

Moreover, choices of $\C$ and $\D$ for other triggering rules from the literature are presented in \cite{postoyan2015framework}.

Further potential of the framework lies in the fact that it can help to generate new triggering rules. In \cite{postoyan2015framework}, this is demonstrated for a triggering rule that includes a hybrid clock. Such hybrid clocks are well known in the analysis of sampled-data systems \citep{nesic2009explicit}. In \cite{postoyan2015framework}, it is illustrated how the dynamic variable is used to represent the hybrid clock. The stability analysis is then carried out within the proposed framework. In \cite{abdelrahim2016stabilization,dolk2017output}, the general concept of the newly found mechanism is even further developed into an ETC triggering rule that leads to a guaranteed $\L_p$ gain for an auxiliary performance channel. 
The framework thus not only allows to unify the analysis of existing triggering rules but instead also to design novel triggering rules with favorable properties. 

A conceptually similar but more general framework is proposed in \cite{chai2020analysis}. Based on hybrid inclusions, a stability analysis is used that allows a Lyapunov function to increase during flows or jumps if this is compensated during jumps or flows, respectively.
This offers the potential for novel designs of triggering rules and may facilitate the analysis of existing ones.
In \cite{heemels2013periodic}, a conceptually related analysis technique via an impulsive systems approach is proposed for PETC and linear time-invariant systems.

While the frameworks from \cite{postoyan2015framework,chai2020analysis}  offer the possibility to analyze a variety of ETC triggering rules, we next introduce a framework based on different technical concepts that is particularly well suited to redesign and improve existing triggering rules. 

\subsection{A framework based on a hybrid small-gain theorem}
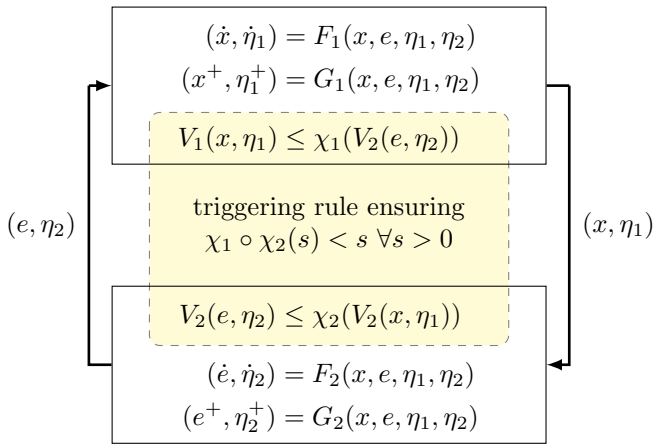
\begin{figure}[th!]
	\centering
	\input{img/fig_framework_hybrid.tex}
	\caption{Block diagram of the feedback interconnection studied in \cite{maass2023event} with effect of the triggering rule on the subsystems highlighted (yellow box).}
		\label{fig_framework_hybrid}
\end{figure}

In the framework from \cite{maass2023event}, again a hybrid system model similar to \eqref{eq_hyb} is considered. 
However, the approach for the analysis of ETC schemes is different from those in \cite{postoyan2015framework,chai2020analysis}.
In particular, instead of Lyapunov conditions, a hybrid small-gain theorem from \cite{liberzon2014lyapunov} is used to analyze properties of the closed-loop system. 
Small gain approaches are also successfully used  in \cite{liberzon2014lyapunov,liu2015small} to analyze the relative threshold triggering rule.
The basic idea in \cite{maass2023event} is to consider the closed-loop system with ETC as a feedback interconnection of two subsystems. 
The first subsystem describes the dynamics of the plant and controller states, i.e., those of $x$.
The input to the first subsystem is the sampling induced error $e$. 
The dynamics of $e$ are described by the second subsystem.
These dynamics also depend on $x$, which is thus an input to the second subsystem within the feedback interconnection.
For dynamic ETC approaches, the dynamic variable $\eta$ is split into two parts $\eta_1$ and $\eta_2$ with dimensions $n_{\eta_1}$ and $n_{\eta_2}$ that are assigned to the respective subsystems.
We can thus rewrite the hybrid system \eqref{eq_hyb} as the two interconnected subsystems
\begin{equation*}
		\begin{cases}
			(\dot{x},
			\dot{\eta}_1) = F_1(\xi) & \xi\in\C\\
			({x}^+,
			{\eta}_1^+) = G_1(\xi) & \xi\in \D
	\end{cases}
\end{equation*}
and 
\begin{equation*}
	\begin{cases}
			(\dot{e},
			\dot{\eta}_2) = F_2(\xi) & \xi\in\C\\
			({e}^+,
			{\eta}_2^+) = G_2(\xi) & \xi\in \D
	\end{cases},
\end{equation*}
where $F_1, F_2, G_1$ and $G_2$ describe the respective parts of $F$ and $G$. 
The hybrid small gain condition \cite[Theorem~2]{maass2023event} essentially requires that there are functions $V_1:\R^{n_x+n_{\eta_1}}\rightarrow\R_{\geq 0}$, $V_2:\R^{n_u+n_{\eta_2}}\rightarrow\R_{\geq 0}$, $\alpha_1,\alpha_2 \in \K_\infty$ and $\chi_1, \chi_2 \in \K_\infty \cup \left\lbrace0\right\rbrace$, such that
\begin{enumerate}[label=(\roman*)]
	\item \label{item_small_gain_i}  The functions $V_1$ and $V_2$ are upper and lower bounded by $\K_\infty$ functions.
	\item \label{item_small_gain_ii} For almost all $\xi\in\C$, 
	\begin{subequations}
	\begin{align}
		\nonumber V_1(x,\eta_1) \geq{}& \chi_1(V_2(e,\eta_2))\\
						\Rightarrow&   V_1^\circ((x,\eta_1); F_1(\xi))  \leq -\alpha_1\left(\abs{(x,\eta_1)}\right) \label{eq_small_gain_a}\\
						\nonumber V_2(e,\eta_2) \geq{}& \chi_2(V_1(x,\eta_1))\\
						\Rightarrow&   V_2^\circ((e,\eta_2); F_2(\xi))  \leq -\alpha_2\left(\abs{(e,\eta_2)}\right). \label{eq_small_gain_b}
	\end{align}
	\end{subequations}
	\item \label{item_small_gain_iii} For all $\xi\in\D$, $V_1(G_1(\xi)) \leq V_1(x,\eta_1)$ and $V_2(G_2(\xi)) \leq V_2(e,\eta_2)$ holds. 
	\item \label{item_small_gain_iv} The small-gain condition $\chi_1\circ\chi_2(s) < s$ holds for all $s > 0$.
	\item There is a non-vanishing time between jumps of the hybrid system \eqref{eq_hyb}.
\end{enumerate}
Conditions~\ref{item_small_gain_i}-\ref{item_small_gain_iii} essentially require that the subsystems are ISS with gains $\chi_1$ and $\chi_2$, respectively.
Item~\ref{item_small_gain_iv} is a small-gain condition for hybrid systems, that is adopted from \cite{liberzon2014lyapunov}. In addition with the ISS properties that are ensured by \ref{item_small_gain_i}-\ref{item_small_gain_iii}, it allows to construct a Lyapunov function and to conclude asymptotic stability. 
To analyze triggering rules, the above conditions are verified depending on the considered triggering rule. 
Recall to that end, that the flow set $\C$ and the jump set $\D$ are determined by the respective triggering rule. 
It turns out, that for many common triggering rules from the literature, \eqref{eq_small_gain_a} actually holds with $\chi_1 = 0$. 
For $\chi_2$, \cite{maass2023event} propose to make a choice ensuring that \eqref{eq_small_gain_b} holds vacuously, i.e., such that $  V_2^\circ((e,\eta_2), F_2(\xi)) \leq -\alpha_2\left(\abs{(e,\eta_2)}\right)$ always holds on $\C$. 
A block diagram of the feedback interconnections with the influence of the triggering rule is given in Figure~\ref{fig_framework_hybrid}.

\subsubsection*{Application of the framework to specific triggering rules}
For the relative threshold triggering rule, the flow and jump sets are given by \eqref{eq_flow_jump_rel}. 
The gains can be obtained as $\chi_1(s) = 0$ and $\chi_2(s) = (1+\varepsilon) \left(\gamma^{-1}\circ\sigma \alpha \circ \underbar{$\alpha$}^{-1}(s)\right)^2$,
where $\varepsilon > 0$ can be chosen arbitrarily. 
It is verified in \cite[Corollary~1]{maass2023event} that this choice indeed satisfies the hybrid small-gain condition and thus asymptotic stability can be concluded. 
Moreover, various different ETC triggering rules are covered by the framework from \cite{maass2023event}, including the absolute threshold triggering rule, the dynamic triggering rule and a triggering rule with a decreasing threshold on a Lyapunov function. 
In \cite{maass2023event}, respective choices for $\chi_1$ and $\chi_2$ are presented for those triggering rules.
This illustrates the versatility of the framework.  

An important aspect of the framework is its potential for redesigning the triggering rules in order to derive potentially less conservative variants. 
Common triggering rules from the literature typically allow the choice $\chi_1 = 0$.
Whilst this trivially verifies Item~\ref{item_small_gain_iv} of the hybrid small-gain condition, it is unnecessarily conservative. 
The aim of the redesign is to reduce the conservatism, typically by increasing the flow set such that the hybrid small-gain conditions still hold with nonzero $\chi_1$.
To illustrate this, we consider again the relative threshold triggering rule. 
By redefining the flow- and jump set to 
\begin{align*}
	\C={}&\left\lbrace \xi\mid (1-\delta)\gamma(\abs{e}) \leq \sigma\alpha(\abs{x}) \right\rbrace,\\ 
	\D ={}& \left\lbrace \xi\mid (1-\delta)\gamma(\abs{e}) \geq \sigma\alpha(\abs{x}) \right\rbrace,
\end{align*}
for sufficiently small $\delta > 0$, the flow set is enlarged whilst the jump set shrinks. 
The redefined sets correspond to the triggering rule 
\begin{equation}
	t_{j+1} = \inf\left\lbrace t > t_j\mid (1-\delta)\gamma(\abs{e(t)}) \geq \sigma \alpha(\abs{x(t)})\right\rbrace.
\end{equation}
Here the term $(1-\delta)$ relaxes the redesigned triggering rule compared to the original variant with the aim to reduce the sampling rate.
For the redesigned triggering rule, the hybrid small gain condition holds if $\delta \in \left(0,\frac{1-\sigma}{1+\sigma+\sigma c}\right)$, where $c$ is a parameter that depends on $\alpha, \underbar{$\alpha$}$ and $\bar{\alpha}$, with  
\begin{align*}
	\chi_1(s) ={}& \bar{\alpha} \circ \alpha^{-1} \circ \frac{\delta}{\nu(1-\sigma)} \gamma\left(s^{1/2}\right),\\
	\chi_2(s) ={}& \left[\gamma^{-1}\circ(1+\varepsilon) \frac{\sigma}{1-\delta} \alpha \circ \underbar{$\alpha$}^{-1}(s)\right]^2,
\end{align*}
where $\varepsilon > 0$ and $\nu\in\left(0,1\right)$, cf.\ \cite[Corollary~2]{maass2023event}. 
Thus, stability guarantees also apply to the redesigned triggering rule.
Redesigns for other triggering rules are studied as well in \cite{maass2023event}.
Moreover, the proposed redesigns in \cite{maass2023event} are not exhaustive and other redesigns are possible as well. 
It is demonstrated in \cite{maass2023event} for a numerical example, that the proposed redesign can lead to triggering rules with larger inter-event times compared to the original ones.
However, there is no guarantee that the redesigns always lead to larger inter-event times. 

There are also triggering rules to which the proposed framework from \cite{maass2023event} is not directly applicable, including those with time regularization \citep{borgers2016dynamic}, event-holding control \citep{wang2019state} and the triggering rule based on a hybrid clock from \cite{postoyan2015framework}.
As concluded in \cite{maass2023event}, it therefore seems sensible to look for even more general frameworks based on dissipativity in the future.
Before discussing this as a perspective for future work, we will first review another framework that takes an alternative small-gain perspective. 

\subsection{A framework based on \texorpdfstring{$\L_p$}{Lp} norms}

The framework in \cite{hertneck2024norm} is inspired by the framework for the analysis of TTC in \cite{nesic2004input}.
Similar to \cite{maass2023event}, the system is modeled as a feedback interconnection of two subsystems:
One subsystem describes the behavior of plant and controller states, and thus of the $x$-dynamics, and the other the behavior of the  network-induced error, and thus of the $e$-dynamics. 
The closed-loop system with ETC can hence be rewritten as the interconnection of the two impulsive subsystems 
\begin{subequations}	 
	\label{eq_sys_gen_2}
	\begin{align}
		\nonumber\dot{x} &= f(x,\kappa(x)+e)\quad t\in\left[t_j,t_{j+1}\right)\\
		y_x&= H(x,e) 	\label{eq_sys_x}\\
		\nonumber x(t_j^+) &= x(t_j)
		\intertext{and}
		\nonumber \dot{e} &= -\frac{\partial \kappa(x)}{\partial x} f(x,\kappa(x)+e) \quad t\in\left[t_j,t_{j+1}\right)\\
		y_e&= W(e)\label{eq_sys_e}\\
		\nonumber e(t_j^+) &= 0.
	\end{align}	
\end{subequations}
Here, $H:\R^{n_x\times n_e} \rightarrow \R^{n_H}$ and $W:\R^{n_e}\rightarrow\R^{n_{W}}$ with $n_H \in \N$, $n_{W} \in\N$, and $W(0) = 0$ determine outputs of the respective system. 

The key difference compared to \cite{maass2023event} is that instead of the hybrid small-gain theorem from \cite{liberzon2014lyapunov}, a classical small-gain theorem based on $\L_p$ norms is leveraged. 
We denote the $\L_p$ norm of a measurable, locally integrable signal $\varphi\colon\left[t_a, t_b\right]\rightarrow \R^n$ for $p\in[1,\infty)$ as 
\begin{equation*}
	\norm{\varphi}_{\L_p\left[t_a,t_b\right]} \coloneqq \left(\int_{t_a}^{t_b} \abs{\varphi(s)}^p ds\right)^{\frac{1}{p}}.
\end{equation*}
For $p = \infty$, we denote the $\L_p$ norm as
\begin{equation*}
	\norm{\varphi}_{\L_\infty\left[t_a,t_b\right]} \coloneqq \ess \underset{s\in\left[t_a,t_b\right]}{\sup}\abs{\varphi(s)}.
\end{equation*}

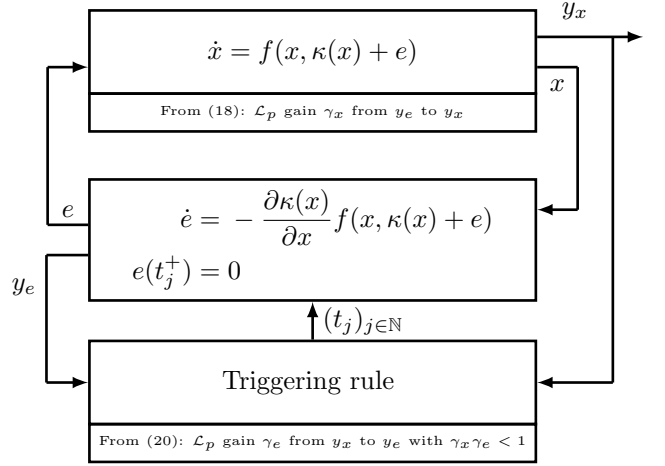
\begin{figure}
	\centering
	\input{img/fig_framework_lp.tex}
	\caption{Block diagram of the feedback interconnection studied in \cite{hertneck2024norm} with effect of the triggering rule on the $e$-subsystem highlighted. }
	\label{fig_framework_lp}
\end{figure}

The first step within the framework from \cite{hertneck2024norm} is to show that the $x$-subsystem \eqref{eq_sys_x} is $\L_p$ stable from input $e$ to output $y_1$ with gain $\gamma_x$, i.e., to determine $\gamma_x>0$ such that the finite-gain $\L_p$ stability condition
\begin{equation}
	\label{eq_cond_x}
	\begin{split}
		&\norm{y_x}_{\L_p[t_0,t]}
		\leq{} K_x(\abs{x(t_0)}) + \gamma_x\norm{y_e}_{\L_p[t_0,t]} 
	\end{split}
\end{equation}
holds for some $K_x \in\K$ and all $t\in\left[t_0,T\right]$.
Then, standard small-gain arguments can be used to show that the interconnection is $\L_p$ stable, if the $e$-subsystem satisfies a similar $\L_p$ stability condition of the form
\begin{equation}
	\label{eq_cond_e}
	\begin{split}
		&\norm{y_e}_{\L_p[t_0,t]}
		\leq{} K_e(\abs{e(t_0)}) + \gamma_e\norm{y_x}_{\L_p[t_0,t]} 
	\end{split}
\end{equation}
for some $K_e \in\K$ with $\gamma_e\gamma_x < 1$, cf.\ \cite[Proposition~1]{hertneck2024norm}. 
The triggering rule needs to be designed such that \eqref{eq_cond_e} holds.
A natural choice is thus
\begin{equation}
	\label{eq_def_gain}
	t_{j+1} = \inf\left\lbrace t > t_j \mid \norm{y_e}_{\L_p\left[t_j,t\right]} \geq    \gamma_e \norm{y_x}_{\L_p\left[t_j,t\right]} \right\rbrace,
\end{equation}
i.e., to trigger transmissions when otherwise the $\L_p$ norm of $y_e$ since the last sampling instant would exceed $\gamma_e$ times the $\L_p$ norm of $y_x$ since the last sampling instant.
A block diagram of the feedback interconnection with this triggering rule is visualized in Figure~\ref{fig_framework_lp}.
Using this type of triggering rule verifies \eqref{eq_cond_e} recursively with equality.
Different specific triggering rules result from using different $p$ norms in \eqref{eq_def_gain}. 
Alternatively, any triggering rule that  for the same $t_j$, $x(t_j)$ and $e(t_j)$ picks $t_{j+1}$ smaller than \eqref{eq_def_gain} for some $p$ can be used in the framework since they ensure as well that \eqref{eq_cond_e} holds.

To show that there is a non-vanishing inter-event time, the approach to determine the maximum inter-event time for TTC from \cite{nesic2004input} can be leveraged.
Alternatively, specific approaches from the literature for specific triggering rules can be used to show that the inter-event time does not vanish. 

\subsubsection*{Application of the framework to specific triggering rules}
As a simple illustration, the relative threshold triggering condition can for example be studied within the framework from \cite{hertneck2024norm} using either $p = 1$ or $p = \infty$. 
In the former case, it is leveraged that the ISS condition Assumption~\ref{asum_iss} implies the $\L_1$-gain condition 
\begin{equation*}
	\norm{\alpha(\abs{x})}_{\L_1\left[t_0,t\right]} \leq \bar{\alpha}(\abs{x(t_0)}) +  \norm{\gamma(\abs{e})}_{\L_1\left[t_0,t\right]}
\end{equation*}
and thus $\L_1$ stability according to \eqref{eq_cond_x} of the $x$-subsystem in \eqref{eq_sys_gen_2} for $H(x,e) = \alpha(\abs{x})$, $W(e) = \gamma(\abs{e})$ and $\gamma_x = 1$. 
Moreover, the relative threshold triggering condition implies \eqref{eq_cond_e} for $p = 1$, $H(x,e) = \alpha(\abs{x})$ and $W(e) = \gamma(\abs{e})$ and is thus covered by the framework. 
Alternatively, for $p = \infty$, it can be shown that Assumption~\ref{asum_iss} also implies
a finite $\L_\infty$ gain according to \eqref{eq_cond_x} (but potentially with a different gain compared to the $\L_1$ case). 
Again, the relative threshold triggering condition implies  \eqref{eq_cond_e} for $p = \infty$. 
Hence, the relative threshold triggering rule can be studied alternatively using the $\L_\infty$-norm condition.
Then different values for $\sigma$ may be admissible compared to the analysis using the $\L_1$ norm.
Thus, more flexibility in the design of the triggering rule can be obtained.
A related ISS condition for the $\L_\infty$-norm case was also used in  \cite{liu2015small} to study the relative threshold triggering rule.

For $\beta = 0$, the dynamic triggering rule \eqref{eq_def_dyn} corresponds exactly to the $\L_1$-norm condition according to \eqref{eq_def_gain} with $\gamma_e = \sigma$, $H(x,e) = \alpha(x)$, and $W(e) = \gamma(e)$ and can thus straightforwardly be studied in the framework.
For other choices of $\beta$, it can further be shown that, for the same $t_j$, $x(t_j)$, and $e(t_j)$, the next sampling instant $t_{j+1}$ is picked smaller than in \eqref{eq_def_gain} with $\gamma_e = \sigma$, $H(x,e) = \alpha(x)$, and $W(e) = \gamma(e)$.
Thus, the dynamic triggering rule can also be studied in the framework and $\L_1$ stability can be concluded since $\sigma \in \left(0,1\right)$.
Asymptotic stability again follows from $\alpha\in \K_\infty$. 

The absolute threshold triggering rule \eqref{eq_trigger_abs} directly implies that the $\L_\infty$-norm bound 
	$\norm{\gamma(e)}_{\L_\infty[t_j,t_{j+1})}\leq d$
holds and thus that 
\begin{equation*}
	\norm{\gamma(e)}_{\L_\infty\left[t_0,t\right]}\leq d
\end{equation*} 
holds for all $t\geq t_0$. 
This can be interpreted as a variant of \eqref{eq_cond_e} with additional constant factor $d$ but with gain $\gamma_e = 0$. 
Such a constant factor can easily be treated within the framework from \cite{hertneck2024norm}. 
For using the absolute threshold triggering rule, it can be used that Assumption~\ref{asum_iss} implies that \eqref{eq_cond_x} holds for some $\gamma_x, W\in\K_\infty$ and $H\in\K_\infty$.
Hence, the absolute triggering threshold can be embedded into the framework using the $\L_\infty$-norm case.

Adding constant terms to \eqref{eq_cond_x} and \eqref{eq_cond_e}  is not only useful to analyze the absolute threshold triggering rule, but also allows studying the robustness of triggering rules from the literature with respect to disturbances.
In particular, if the $\L_p$ gain from a disturbance input to $y_x$ is finite for the closed-loop system with continuous feedback (i.e., for the $x$-subsystem if $e = 0$) then the triggering rule can be modified such that the $\L_p$ gain for the system with ETC is also finite. 

Similar as for the framework from \cite{maass2023event}, there are however some triggering rules for which it is not clear how they can be covered by the framework from \cite{hertneck2024norm}, including the one based on a hybrid clock from \cite{postoyan2015framework}. 

\subsection{Perspectives for further research}
The frameworks presented above provide unified stability proofs and interpretations for large classes of ETC triggering rules and therefore contribute to the understanding of ETC in general. 
However, there is still potential for more general frameworks. 
In particular, it seems a natural next step to aim for more general frameworks based on dissipativity theory  \citep{willems2007dissipative}.
Dissipativity can be interpreted as a generalization of Lyapunov theory.
It also covers small-gain results and passivity as a special case.
It therefore has the potential to subsume frameworks presented in this section and, thereby, can lead to a unified perspective on ETC analysis, shedding light on the relationship between currently coexisting frameworks.

Following the approach from \cite{maass2023event,hertneck2024norm}, there is great potential in interpreting the system with ETC as an interconnection of two dissipative systems, one describing the behavior of plant and controller and one describing the effects of the ETC triggering rule. 
If supply rates are known for both subsystems, analysis tools such as those from \cite{arcak2016networks} can  be used. 
Given that supply rates for specific triggering rules are known, properties of the closed-loop system could be determined. 
These properties could go beyond pure stability considerations and could also cover dissipativity properties for the closed-loop system.
In addition, robustness with respect to disturbances and model uncertainties can naturally be embedded into a dissipativity framework. 
Moreover, seeking for triggering rules that satisfy specific dissipation properties can lead to new triggering rules that directly verify dissipation inequalities or, alternatively, imply that dissipation properties hold. 
Besides determining possible supply rates that result for specific triggering rules, another open problem for such a framework is to develop unified techniques to guarantee that the inter-event time does not vanish. 

Also related to a dissipativity perspective, another approach to find generalized interpretations for ETC triggering rules is to take a robust control perspective. 
In such a setup, the interpretation of the ETC triggering rule is that it ensures that the subsystem describing the sampling induced error satisfies a chosen uncertainty description. 
Then, robust control techniques can be used for the stability analysis of the closed-loop system. 
Such an uncertainty approach was, e.g., pursued in \cite{hertneck2024event}, where transmissions are triggered such that an integral quadratic constraint (IQC) for the subsystem that describes the sampling-induced error is guaranteed to hold. 
The IQC-framework from \cite{veenman2016robust} is then used for the stability analysis.
While the approach of \cite{hertneck2024event} is limited to a specific triggering rule, it seems promising to use a robust control perspective also for different triggering rules. 
General frameworks based on robust control theory not only bear the potential to better understand effects in ETC but also offer powerful analysis tools for various system properties and even offer potentials for controller synthesis.

Another promising research direction is the development of unifying frameworks for STC. 
Such frameworks could aim to cover large classes of ETC triggering rules to approximate them with STC techniques. 
STC techniques developed for such frameworks could then be applied to the covered triggering rules. 
This could provide a modular toolbox for combining STC techniques and ETC triggering rules, allowing suitable combinations to be used depending on the application.

%% file: img/fig_framework_hybrid.tex
	\begin{tikzpicture}[ >=latex]
	\pgfdeclarelayer{background}
	\pgfdeclarelayer{foreground}
	\pgfsetlayers{background,main,foreground}
	
	\node[draw, minimum height=.8cm,anchor=south west,minimum height=1cm] at (0,0) (state){\begin{minipage}{5.5cm}
			\vspace{-4mm}
			\begin{align*}
				(\dot{x},\dot{\eta}_1) ={}& F_1(x,e,\eta_1,\eta_2)\\
				(x^+,\eta_1^+) ={}& G_1(x,e,\eta_1,\eta_2)\\
				\\[-.8em]
				V_1(x,\eta_1) \leq{}& \chi_1 (V_2(e,\eta_2))
			\end{align*}
	\end{minipage}};
	\node[draw,below= 16mm of state,minimum height=1.5cm,align=center] (error) {\begin{minipage}{5.5cm}
			\vspace{-4mm}
			\begin{align*}
				V_2(e,\eta_2) \leq{}& \chi_2 (V_2(x,\eta_1))\\
				\\[-.8em]
				(\dot{e},\dot{\eta}_2) ={}& F_2(x,e,\eta_1,\eta_2)\\
				(e^+,\eta_2^+) ={}& G_2(x,e,\eta_1,\eta_2)
			\end{align*}
	\end{minipage}};
	
	\node (trigger) at ($(state)!0.5!(error)$) {
		\begin{minipage}{4.5 cm}
			\centering
			triggering rule ensuring\\ $\chi_1\circ\chi_2(s) < s~\forall s> 0$
	\end{minipage}};
	
	\coordinate[right = 3 mm of state](state_right);
	\coordinate[left = 3 mm of state](state_left);
	
	\draw [-,line width = 1pt]  (state.east) -- (state_right);
	\draw [<-,line width = 1pt]  (state.west) -- (state_left);
	
	\draw [->,line width = 1pt] (state_right) |- node [pos = 0.25,right = 0.1em,align = center] {$(x,\eta_1)$}(error.east);
	\draw [-,line width = 1pt] (state_left) |- node [pos = 0.25,left = 0.1em,align = center] {$(e,\eta_2)$}(error.west);
	
	\begin{pgfonlayer}{background}
		\path (state.west |- state.north)+(.5,-1.4) node (a) {};
		\path (error.south -| error.east)+(-.5,1.3) node (b) {};
		\path[fill=yellow!20,rounded corners, draw=black!50, dashed]
		(a) rectangle (b);
	\end{pgfonlayer}
	
\end{tikzpicture}

%% file: img/fig_framework_lp.tex
\pgfdeclarelayer{background}
\pgfdeclarelayer{foreground}
\pgfsetlayers{background,main,foreground}
\begin{tikzpicture}
	[line width=1.0, >=latex]
	\usetikzlibrary{calc}
	\node[draw,minimum width=5.9cm,minimum height=.8cm,anchor=south west,minimum height=1.1cm] at (0,0) (node1){\begin{minipage}{3.3cm}\centering$\dot{x} = f(x,\kappa(x)+e)$\end{minipage}};
	\node[draw,below= 11mm of node1, minimum width=5.9cm,minimum height=1.6cm,align=center] (node2) {\begin{minipage}{5cm}
			\centering	
			\vspace{-4mm}\begin{align*}
				\dot{e} ={}& - \frac{\partial \kappa(x)}{\partial x} f(x,\kappa(x)+e)\\
				e(t_j^+) ={}& 0
			\end{align*}
	\end{minipage}};
	\node[draw,below= 5mm of node2, minimum width=5.9cm,minimum height=1.1cm,align=center] (node3) {
		\begin{minipage}{3.3cm}
			\centering Triggering rule
	\end{minipage} };

	\node[draw,below= -1pt of node1, minimum width=5.9cm,align=center,minimum height= 5mm] (node4) {\tiny From \eqref{eq_cond_x}: $\L_p$ gain {\color{black}$\gamma_x$} from $y_e$ to $y_x$};
	
	\node[draw,below= -1pt of node3, minimum width=5.9cm,align=center,minimum height= 5mm] (node5) {\tiny From \eqref{eq_def_gain}: $\L_p$ gain {\color{black}$\gamma_e$} from $y_x$ to $y_e$ with $\gamma_x\gamma_e < 1$};

	\coordinate[left = 0 mm of node1](Q0a);
	\coordinate[below = 2 mm of Q0a](Q0b);
	\coordinate[left = 0 mm of node2](Q1a);
	\coordinate[above = 0 mm of Q1a](Q1b);
	\coordinate[below = 0 mm of Q1a](Q1c);
	
	\coordinate[right = 0 mm of node1](Q2a);
	\coordinate[below = 2 mm of Q2a](Q2b);
	\coordinate[above = 2 mm of Q2a](Q2c);
	
	\coordinate[right = 0 mm of node2](Q3a);
	\coordinate[above = 4 mm of Q3a](Q3b);

	\coordinate[below = 0 mm of node1](Q4);
	\coordinate[left = 3 mm of Q4](Q4a);
	\coordinate[right = 3 mm of Q4](Q4b);
	
	\coordinate[above = 0 mm of node2](Q5);
	\coordinate[left = 3 mm of Q5](Q5a);
	\coordinate[right = 3 mm of Q5](Q5b);
	\node[left = 4 mm of Q0b] (helpnode0) {};
	\node[left = 4 mm of Q1b] (helpnode1) {};
	\node[right = 4 mm of Q2b] (helpnode2) {};
	\node[right = 4 mm of Q3b] (helpnode3) {};
	\coordinate (Q0) at (helpnode0);
	\coordinate (Q1) at (helpnode1);
	\coordinate (Q2) at (helpnode2);
	\coordinate (Q3) at (helpnode3);
	\coordinate[above = 2mm of Q0a] (Q15); 
	\coordinate[left = 5mm of Q15] (Q16); 
	
	\coordinate[below = 4mm of Q3a] (Q17); 
	\coordinate[right = 5mm of Q17] (Q18);
	
	\coordinate[above = 2mm of Q1b] (Q1d);
	\coordinate[above = 2mm of Q1] (Q1e);
	\draw (Q0) to[-,line width = 1pt] (Q1e);
	\draw (Q2b) [-,line width = 1pt] -- node [,midway,,below =0.1em,align = center]{$x$} (Q2);
	\draw [->,line width = 1pt] (Q3) to (Q3b);
	\draw (Q3) to[-,line width = 1pt]  (Q2);
	
	\draw [-,line width = 1pt] (Q1e) -- node [,midway,above =-0.1em,align = center] {$e$}(Q1d);
	\draw [->,line width = 1pt] (Q0) -- node [,midway,above = 0.0em,align = center] {} (Q0b);

	\coordinate[right = 14 mm of Q2c](Q2ca);
	\coordinate[right = 10 mm of Q2c](Q2cb);
	\coordinate[right = 2mm of Q2c] (Q2cc);
	\draw [->,line width = 1pt, color=black] (Q2c) -- (Q2ca);
	\path [line width = 1pt, color=black] (Q2cc) -- node [align = left,above = 1mm,near start]{$y_x$}(Q2ca);

	\coordinate[left = 5.5 mm of node3](Q20);
	\coordinate[right = 10 mm of node3](Q21);
	
	\draw [->,line width = 1pt, color=black] (Q20) -- (node3);
	\draw [->,line width = 1pt, color=black] (Q21) -- (node3);
	\draw [-,line width = 1pt, color=black] (Q21) -- (Q2cb);
	
	\coordinate[below = 2mm of Q1c] (Q1f);
	\coordinate[left = 5.5 mm of Q1f](Q1ca);
	\draw [-,line width = 1pt, color=black] (Q1ca) -- (Q1f);
	\draw [-,line width = 1pt, color=black] (Q1ca) -- node [align = right,left = .0 cm,near start]{$y_e$} (Q20);
	\coordinate[left = 7.5mm of Q20] (Q20b);
	\path [-,line width = 1pt,color = black] (Q20b) -- (node3);
	
	\draw [->,line width = 1pt, color=black] (node3) -- node [align = center,right = 0cm]{$(t_j)_{j\in\mathbb{N}}$}(node2);

\end{tikzpicture}

%% file: sampling_behavior.tex
\section{Analyzing the sampling behavior of event-based control}
\label{sec_sampling}

In the previous sections, we have presented different approaches to design triggering rules for ETC. 
We have discussed for specific triggering rules as well as for unifying frameworks, how guarantees for stability and a lower bound for the inter-event time can be obtained. 
These are also the guarantees that are typically provided for other event-based control schemes in the literature. 
It is however important to analyze the sampling behavior beyond a lower bound for the inter-event times, e.g., understanding what (average) inter-event times  can be expected.
This is not only motivated by the need for a deeper theoretical understanding of event-based control designs, but also by practical reasons.
For practical applications, it is important to know which resources are required when using event-based control. 

To further illustrate why such analyses are beneficial, we consider a simple example system with
\begin{equation}
	\label{eq_sys_sample_ex}
	\dot{x} = \begin{bmatrix}
		0 & 1\\
		-2 & 3
	\end{bmatrix} x
	+ \begin{bmatrix}
		0 & 0\\
		-1 & 4
	\end{bmatrix} \hat{u},
\end{equation}
where $x(t)\in\R^2$ and $u(t)\in\R^2$, and the state-feedback controller
\begin{equation}
	\label{eq_contr_sample_ex}
	u = x.
\end{equation}  
We consider the relative threshold ETC triggering rule 
\begin{equation}
	t_{j+1} = \inf\left\lbrace t > t_j\mid \abs{e(t)}^2 \geq \sigma \abs{x(t)}^2\right\rbrace,
	\label{eq_trigger_rel_sample}
\end{equation}
where $e = \hat{u}-u$. 
\begin{figure}
	\centering
	\input{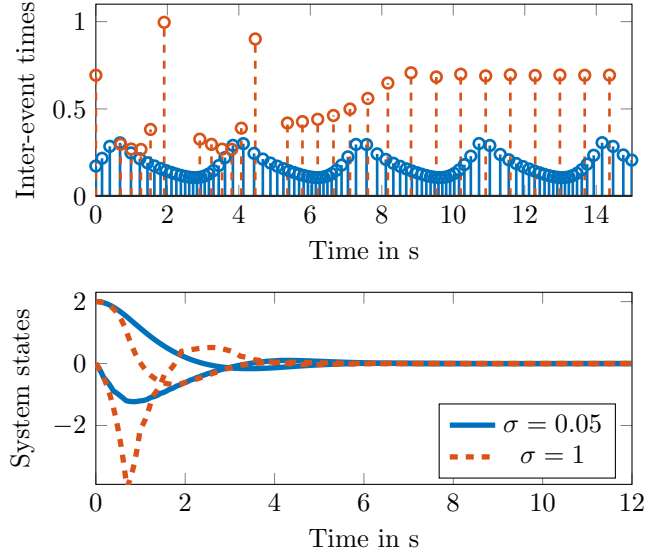}
	\caption{Visualization of trajectories (below) and inter-event times (above) for simulations of the example system \eqref{eq_sys_sample_ex}, \eqref{eq_contr_sample_ex} with triggering rule \eqref{eq_trigger_rel_sample} with $\sigma = 0.05$ and $\sigma = 1$ with initial condition $x(0) = (2,0)$.}
	\label{fig_sampling_example}
\end{figure}
Plots of simulations for this example system with different choices for $\sigma$ and the same initial condition $x(0) = (2,0)$ are given in Figure~\ref{fig_sampling_example}.
In the plot for the inter-event times, it can be seen that for $\sigma = 0.05$, the inter-event times fluctuate significantly but nevertheless follow a periodic pattern.
For $\sigma = 1$, there is first a transient phase in which the inter-event times are varying significantly, and then they eventually converge. 
The different threshold parameters also lead to significantly different state trajectories. 
This simulation example has multiple implications. 
Firstly, we see that the sampling patterns can heavily depend on chosen parameters like the threshold parameter $\sigma$.
Note that sampling patterns can also be chaotic, as demonstrated for example in \cite{gleizer2023chaos} with only slight variations to the system dynamics \eqref{eq_sys_sample_ex}.
Secondly, there can be different phases within one simulation.

Even if the inter-event times vary significantly for some time, it can still happen that they eventually converge. 
To understand which resources are needed for systems with event-based control, it is thus important to understand when which patterns occur.
For that, comprehensive tools for the analysis of the sampling behavior of event-based control that go beyond lower bounds on the inter-event times are required. 
In this section, we provide an overview over such analysis tools.

For simplicity, we restrict ourselves in this section to linear systems with $f(x,\hat{u}) = Ax+B\hat{u}$ and $\kappa(x) = K x$ for matrices $A, B$, and $K$ of suitable dimensions and a relative threshold triggering rule with quadratic functions $\alpha(s) = \gamma(s) = s^2$. 
This makes it, similar as in Section~\ref{sec_stc}, possible to explicitly describe state trajectories. 
For $\delta\in\left(0,t_{j+1}-t_j\right)$, it holds that
\begin{equation*}
	x(t_j+\delta) = G(\delta) x(t_j)
\end{equation*}
where, similar as in Section~\ref{sec_stc}, 
\begin{equation*}
	 G(\delta) \coloneqq e^{A \delta}  + \int_{0}^{\delta} e^{A(\delta-\tau)} BK\mathrm{d}\tau.
\end{equation*}

For analysis purposes, we define $\vartheta:\R^{n_x}\rightarrow \R_{\geq 0}$ as the function that determines the next inter-event time at a sampling instant based on the state $x(t_j)$,
 i.e., the function that satisfies  $ \vartheta(x(t_j)) =  h_j = t_{j+1} - t_j$.
Using the relative threshold triggering rule \eqref{eq_trigger_rel} with design parameter $\sigma$ for the considered specific choice of functions $\alpha$ and $\gamma$, we can write
\begin{equation}
	\label{eq_trigger_ana}
		\vartheta(x(t_j)) \coloneqq \min\left\lbrace \delta > 0\mid x^\top(t_j) M(\delta) x(t_j)  = 0 \right\rbrace,
\end{equation}
where
\begin{equation*}
	M(\delta) \coloneqq (1-\sigma) G^\top(\delta)G(\delta) - \left(G^\top(\delta) + G(\delta)\right) + I.
\end{equation*}
In the general case, it is still difficult to compute $\vartheta(x(t_j))$, as this requires to integrate the matrix exponential function. 
A first order Taylor approximation can be used to derive the following approximate expression. 
\begin{prop}\cite[Proposition~3]{postoyan2023explaining}.
	\label{prop_sampl}
	There exists $r:\R^{n_x}\times\left(0,1\right)\rightarrow\R$, $c_r>0$ and $\sigma^\star\in\left(0,1\right]$ such that for any $\sigma\in\left(0,\sigma^*\right)$ and any $x(t_j)\in\R^{n_x}$, $\vartheta(x(t_j)) = \sigma \frac{\abs{x(t_j)}}{\abs{(A+BK) x(t_j)}} + r(x(t_j),\sigma)$ and $\abs{r(x(t_j),\sigma)} \leq c_r \sigma^2$ holds. 
\end{prop}
Using Proposition~\ref{prop_sampl}, the inter-event time can thus be approximated as  $\vartheta(x(t_j)) \approx \sigma \frac{\abs{x(t_j)}}{\abs{(A+BK) x(t_j)}}$ with an approximation error $r(x(t_j),\sigma)$ quadratically bounded in $\sigma$. 
To analyze sampling patterns, it does however not suffice to approximately compute the value of $\vartheta(x(t_j))$ for specific states, but it is instead important to investigate how inter-event times evolve along trajectories of the system. 
To that end, another important conclusion that can be drawn from \eqref{eq_trigger_ana} is, that inter-event times are constant along rays that pass through the origin. 
Thus, $\vartheta(c x(t_j))$ is constant for varying $c > 0$ if $x(t_j)\in\R^{n_x}$ is fixed. 
The analysis of sampling patterns can hence be reduced to investigating on which rays states at subsequent sampling instants along a trajectory are allocated.  
Whilst this is in general still difficult, approaches for approximations exist under additional assumptions.

We will next discuss two different approaches to do so. 
For planar systems (i.e., systems of dimension $n_x=2$), analytical statements about the sampling behavior are possible under specific conditions. 
Alternatively, the state space can be divided into cones for which upper and lower bounds of the inter-event times can be computed numerically. 
The sampling behavior can then be analyzed based on finite state abstractions that relate the different regions. 
 Subsequently, we summarize both approaches and give an overview of the conclusions that can be drawn.

\subsection{Planar systems}

For planar systems, more specific statements about the resulting sampling instants can be made depending on the eigenvalues. 
We denote by $\lambda_1$ and $\lambda_2$ the eigenvalues of $A+BK$. 
For real, negative eigenvalues, i.e., if $0>\lambda_1>\lambda_2$, we can state the following result.
\begin{theo}\cite[Theorem~3]{postoyan2023explaining}.
	\label{theo_distinct}
	When $0>\lambda_1 > \lambda_2$, there exist $c_1,c_2 > 0$ and $\sigma^\star\in(0,1]$ such that for any initial condition $x(t_0)\in\R^{2}$ and any $\sigma\in(0,\sigma^\star)$, the solution $x(t)$ satisfies one of the following properties:
	\begin{enumerate}
		\item $\underset{j\rightarrow \infty}{\limsup} \abs{\vartheta(x(t_j))-\frac{\sigma}{\abs{\lambda_1}}} \leq c_1 \sigma^2$.
		\item \label{item_distinct_2}  $\abs{\vartheta(x(t_j)) - \frac{\sigma}{\abs{\lambda_2}} }\leq c_2 \sigma^2$ for all $j\geq 0$. 
	\end{enumerate}
\end{theo}
The proof of Theorem~\ref{theo_distinct} exploits that for small values of $\sigma$, the closed-loop system with event-based control behaves approximately like the system $\dot{x}_c = (A+BK) x_c$. 
If the conditions of the Theorem hold, then the state $x(t)$ converges either to the ray  associated to the eigenvector for $\lambda_1$ or it evolves constantly on the ray associated to the eigenvector for $\lambda_2$. 
For either case, the values to which $\vartheta(x(t_j))$ converges can be computed approximately with a remainder that is bounded quadratically in $\sigma$. 
If $\lambda_1 = \lambda_2$ with geometric multiplicity 2,  then it can even be shown, that always the second item holds, cf.\ \cite[Theorem~4]{postoyan2023explaining}. 
A more complicated behavior can arise for systems in which the eigenvalues are complex conjugates of the form $\lambda_1 = \lambda + i\beta$, $\lambda_2 = \lambda - i\beta$ for $\lambda < 0$ and $\beta > 0$. 
Then, the inter-event times may exhibit an oscillatory behavior.
Formally, we can state the following result. 
\begin{theo}\cite[Theorem~1]{postoyan2023explaining}.
	\label{theo_complex}
	When $\lambda_1$ and $\lambda_2$ are non-real complex conjugates with negative real part, then there exist $c_3 >0$, $c_{\mathrm{complex}} > 0$, and $\sigma^\star\in(0,1]$ such that for any initial condition $x(t_0)\in\R^2$ and any $\sigma\in\left(0,\sigma^\star\right)$, the solution $x(t)$ satisfies the following property. 
	For any $j\in\N_0$, there exist
	 $\varkappa(t)\in\left[\frac{\pi}{\beta}- c_{\mathrm{complex}}\sigma, \frac{\pi}{\beta}+ c_{\mathrm{complex}}\sigma\right]$ and $r_{\mathrm{complex}}(t,x(t_0),\sigma)$ such that $\vartheta(x(t_j)) = \vartheta(x (t_j+\varkappa(t_j))) + r_{\mathrm{complex}}(t,x(t_0),\sigma)$ and $\abs{ r_{\mathrm{complex}}(t,x(t_0),\sigma)} \leq c_3\sigma^2$. 
\end{theo}
For $\sigma$ sufficiently small, Theorem~\ref{theo_complex} thus implies that for systems with complex conjugate eigenvalues, inter-event times are approximately periodic along solutions.
It thus also applies to system \eqref{eq_sys_sample_ex}, \eqref{eq_contr_sample_ex} with triggering rule \eqref{eq_trigger_rel_sample},  for which the eigenvalues of $A+BK$ are calculated as $\lambda_{1/2} = -\frac{1}{2} \pm \frac{\sqrt{3}}{2}\mathrm{i}$, if the threshold parameter $\sigma$ is sufficiently small.
This explains the periodic behavior for $\sigma = 0.05$, while $\sigma = 1$ seems to be too large to make a statement based on the theorem.
The proof of Theorem~\ref{theo_complex} again exploits the system dimension and involves showing that solutions are spiraling towards the origin with approximately constant angular velocity.
Then, it can be used that inter-event times on rays passing through the origin are constant.
It can further be shown that the initial condition $x(t_0)$ has only minor impact on the actual sequence of inter-event times, cf.\ \cite[Theorem~2]{postoyan2023explaining}. 
An important conclusion for the case with complex conjugated eigenvalues with imaginary part $\beta$  is thus that a single simulation over $\frac{\pi}{\beta}$ units of time can be used to approximate the inter-event times for all initial conditions and future sampling instants.
The simulation thus leads to an estimate for the average sampling intervals in this case. 
Theorems~\ref{theo_distinct} and \ref{theo_complex}  provide an explanation as to why systems with event-based control often tend to exhibit asymptotically periodic sampling behavior in simulations. 
A drawback of Theorems~\ref{theo_distinct} and \ref{theo_complex} is however, that the conclusions are typically only valid for $\sigma^*$ close to 0.
Then the system with event-based control has trajectories close to those of the respective system with continuous feedback.

An alternative way to approach the analysis of inter-event times for planar systems is to parameterize the state in angular coordinates with $ x(t_j) = c_x(t_j) \begin{bmatrix}
	\cos(\phi_x(t_j)) & \sin(\phi_x(t_j))
\end{bmatrix}^\top$
for $c_x(t_j)> 0$ and $\phi_x(t_j)\in\left[-\pi,\pi\right)$ and directly study the evolution of the angle $\phi_x$ along trajectories. 
A sampling instant is triggered at time $t_j+\delta$ if  $x^\top(t_j) M(\delta) x(t_j)  = 0$. 
Since inter-event times are constant along rays passing through the origin, this is equivalent to 
		$\begin{bmatrix}
			\cos(\phi_x(t_j)) & \sin(\phi_x(t_j))
		\end{bmatrix} M(\delta) \begin{bmatrix}
			\cos(\phi_x(t_j)) \\ \sin(\phi_x(t_j))
		\end{bmatrix}
		  = 0.$
By analyzing properties of $M$, it can be shown, that $\vartheta(x(t_j))$ is purely determined by the angle $\phi_x(t_j)$, cf.\ \cite[Corollary~4]{rajan2024asymptotic}.
We will thus write subsequently equivalently $\vartheta(\phi_x(t_j))$.
The evolution of $\phi_x$ is determined by the angle map $\phi_x(t_{j+1}) = \Psi(\phi_x(t_j)) \coloneqq \arg \left(G(\vartheta(\phi_x(t_j))) \begin{bmatrix}
	\cos(\phi_x(t_j))\\
	\sin(\phi_x(t_j))
\end{bmatrix}\right)$, where $\arg(s)$ denotes the angle between $x(s)$ and the first coordinate axis.
The following statements can be made.
\begin{theo}\cite[Remark~11]{rajan2024asymptotic}.
	Suppose there exists a fixed point of the angle mapping $\Psi$, i.e., $\exists \phi \in[0,\pi)$ s.t. $\Psi(\phi) = \phi$.
	Then, for all initial conditions $x(t_0) = c_x(t_0) \begin{bmatrix}
		\cos(\phi_x(0)) & \sin(\phi_x(0))
	\end{bmatrix}^\top$ with $\phi_x(0) = \phi$, it holds that  $t_{j+1}- t_j = \vartheta(\phi_x(t_j)) = \vartheta(\phi)$ for all $j\in\N_0$. Moreover, if $\phi$ is an asymptotically stable fixed point of the angle map, then $\lim\limits_{j\rightarrow \infty} (t_{j+1}-t_j) = \vartheta(\phi)$ for all initial conditions in the region of convergence of $\phi$ under the angle map $\Psi(\cdot)$. 
\end{theo}
\begin{theo}\cite[Theorem~12]{rajan2024asymptotic}.
	If there does not exist $\phi\in[0,\pi)$ such that $\Psi^k(\phi) - \phi = d\pi$ for $d\in\Z, \forall k\in\left\lbrace1,2\right\rbrace$ and if the inter-event time function $\vartheta(\cdot)$ is not a constant function, then the inter-event times do not converge to a steady state value for any initial condition. 
\end{theo}

Thus, the asymptotic behavior of systems with event-based control can be analyzed by analyzing fixed points of the angle mapping $\Psi$. 
It is however in general still difficult to analyze the angle map. 
Results for specific conditions that facilitate the analysis of the angle map are given in \cite{rajan2024asymptotic}. 
The consideration of fixed points can also be conceptually applied to higher dimensional systems. 
In \cite[Section IV]{gleizer2023chaos}, conditions for the existence and attractivity of invariant hyperplanes for event-based control are presented, which allow drawing conclusion about the asymptotic sampling behavior of higher dimensional systems.
The analysis of periodic sampling patterns using invariance arguments is further studied in \cite{velasco2009equilibrium}.
In \cite{Rajan2024}, convergence of inter-event times to constant or periodic sampling patterns for linear time-invariant systems under region-based STC is studied with the help of positively invariant subregions.

The above approaches provide insightful starting points for better understanding the sampling behavior of systems with event-based control. However, the statements are limited to specific setups and often only cover the asymptotic behavior.
Therefore, it is currently challenging to leverage these results for practical reasoning about required resource capacities, apart from asymptotic considerations.
For this reason, further research is needed for more general setups.

\subsection{Analysis based on finite state abstractions}

An alternative to the analytical approach for the analysis of the sampling behavior is to analyze sampling patterns numerically based on finite state abstractions. 
To that end, it is proposed in \cite{sharifikolarijani2018formal} to construct a finite automaton that abstracts the sampling behavior of a system with event-based control. 
The state space is divided into $n_s\in\N$ convex conic regions of the form\footnote{Note that for systems with $n_x = 2$, a different state space partitioning is used, cf.\ \citep{sharifikolarijani2018formal}.}
\begin{equation*}
	\mathcal{R}_s = \left\lbrace x \in \R^{n_x} \mid E_s x \geq 0 \right\rbrace
\end{equation*}
for $s\in\left\lbrace 1,\dots, n_s\right\rbrace$, $E_s\in\R^{p \times n_x}$ with $p\leq 2n_x-2$.
Exploiting the fact that $\vartheta(x)$ is constant along rays through the origin, upper and lower bounds for the inter-event times are computed  for each region. 
More specifically, for each region, bounds $\underbar{$h$}_s$ and $\bar{h}_s$ are determined such that $x^\top M(\delta) x \leq 0 \;\forall \delta \in [0, \underbar{$h$}_s]$ and $x^\top M(\delta) x \geq 0 \;\forall \delta \in [\bar{h}_s,\infty)$.
This can be done in terms of linear matrix inequalities, cf.\ \cite[Theorems~1 and 2]{sharifikolarijani2018formal}. 
Then, for each region, a reachability analysis is carried out to derive possible transitions to other regions. 
A finite automaton is constructed where the states represent the conic regions. 
Transitions from one state to another are added, if the region of the target state can be reached from the region of the initial state within the bounds for the inter-event time associated to the initial state.
The automaton outputs for each region a time interval of possible values for the inter-event time based on the derived upper and lower bounds. 
In \cite[Theorem~3]{sharifikolarijani2018formal}, it is shown that the above construction approximately simulates the behavior of the true (infinite state) event-based control system up to an error that can be quantified. 
Thus, using the abstraction, possible sampling patterns can be identified. 
The abstraction can, e.g., be used for scheduling \citep{mazo2018abstracted}. 

The accuracy of the abstraction can be adjusted by varying the number of regions. 
In general, a dense partition of the state space into cones leads however to a rapid increase of the computational complexity for higher dimensional systems. 
As an alternative, the regions can be selected based on the inter-event time, i.e., regions are defined as sets of states for which the next inter-event time takes values in the same interval. 
Such partitions based on inter-event times are also used in \cite{delimpaltadakis2023abstracting}, where the above approach is extended to homogeneous nonlinear systems and, using a homogenization procedure, to general nonlinear systems. In this nonlinear setup, also a more complex reachability analysis is required. 

The analysis based on finite state abstractions is also used in \cite{gleizer2021towards,gleizer2023chaos,gleizer2023computing} to determine qualitative properties of linear (periodic) ETC systems using finite state abstractions. 
Therein, approaches to compute different metrics that capture average inter-event times as well as to determine whether chaotic behavior occurs are presented. 
It is even demonstrated in \cite{gleizer2023chaos} that, in specific situations, ETC can lead to less frequent sampling compared to periodic sampling with the largest fixed stabilizing sampling period. 
Moreover, an interesting insight from \cite{gleizer2023chaos} is that complex sampling patterns particularly occur for more ``aggressive'' triggering mechanisms for which trajectories deviate from the trajectories of the systems with continuous feedback. 
This is in line with Theorems~\ref{theo_distinct} and \ref{theo_complex}, which are valid for triggering parameters that ensure that the event-based control system behaves similarly to the system with continuous feedback. 
Probabilistic metrics for stochastic ETC are determined using finite state abstractions in \cite{delimpaltadakis2024formal}.

\subsection{Perspectives for further research}

There are different approaches to characterize the sampling behavior of event-based control systems.
Analytic approaches allow making statements about the asymptotic behavior of linear systems in specific setups.
The restriction to specific setups however limits the applicability of analytic approaches.
To this end, particularly the analysis for systems of dimensions higher than 2 or of more complex setups like systems with output feedback can be considered as a potential goal for future research.

Existing numeric approaches have the advantage, that they can be used for more complex system classes.
Such approaches can also be leveraged for scheduling, cf.\ \cite{gleizer2020scalable}.
In addition, they could also be used in the future to increase average inter-event times. 
To that end, it is noted in \cite{gleizer2021self} that (periodic) ETC is a greedy optimization strategy in the sense that always the short-term reward in the form of the next sampling instant is maximized, hoping that this brings long-term maximization. 
However, triggering at some specific sampling instants earlier than required to obtain stability guarantees may lead to larger inter-event times on average.
Whilst existing work exploits this finding for STC based on abstractions in \cite{gleizer2021self} and using a hierarchical strategy with reinforcement learning in \cite{ong2024hierarchical}, there is still a need for better understanding of the underlying effects in order to find improved ETC schemes.
This knowledge can then also be used in setups with shared resources, like in NCS, in order to design sampling schemes that truely consider long-term resource usage.
For example, knowledge about the expected sampling pattern offers benefits for scheduling and, thus, efficient resource allocation.
More approaches to address the opportunistic nature of ETC in shared resource setups are provided in Section~\ref{sec_current}.
A potential drawback of  numeric approaches that should be addressed by future research is their computational complexity that makes the analysis difficult for higher order systems.

Overall, the results regarding the sampling behavior are quite diverse, depending on the considered setup. 
However, this does not imply that the results are contradictory. 
It rather emphasizes that different setups lead to different properties of the closed-loop system with event-based control.
Thus, a thorough analysis is needed for each specific setup.
To enable such an analysis, further tools need to be developed for setups that are not yet captured by the literature as, e.g., output feedback, dynamic triggering rules, and nonlinear systems. 
An overarching goal for the future is therefore to find a comprehensive mathematical perspective that makes it possible to capture all the observed effects in a unified manner. 

So far, we have focused on the control objective of stabilization of the closed loop while ideally reducing the average sampling rate or increasing the minimum inter-event times with event-based control.
In the next section, we instead study control performance as the respective control objective and, hence, examine the performance sampling rate trade-off.

%% file: performance_comparison.tex
\section{Performance sampling rate trade-off}
\label{sec_comparison}

After having discussed sampling behavior analysis for ETC, let us return to a core motivation for developing and applying event-based control schemes.
As laid out in the introduction, event-based control has the potential to reduce the sampling rate compared to periodic control while still fulfilling control performance requirements, e.g., stability or quantitative performance requirements such as a prescribed LQR cost. %
The seminal work by \cite{astrom1999comparison} proves this advantage of ETC for a first-order stochastic system using an absolute threshold triggering rule and considering the resulting state variance as performance measure.
As for example pointed out in \cite{heemels2012introduction}, the performance comparison between event-based and periodic control schemes is a key enabler of the field and requires both, the consideration of control cost and sampling cost.
Practical examples for the sampling cost are the average rate at which packets are sent over a network or at which computationally expensive operations are performed.
However, as we have seen in the previous section, even the analysis of the sampling behavior and hence any potential sampling cost, e.g., an average sampling rate, remains challenging.
Consequently, this important pillar still offers pressing research questions until today.
In this section, we provide an overview of existing approaches and insights into the performance sampling rate trade-off in event-based control and its relationship to periodic control.
Furthermore, we elaborate on related open research questions and their importance for the field.

\subsection{Optimal design of event-based control}

One line of research has focused on finding optimal design methods for ETC as an approach to navigate the typical trade-off between control performance and sampling cost.
We will explore different formulations for this optimal design problem in this section.
Note that this problem is hard to solve in general, for example, due to the hybrid nature of the closed loop when introducing an ETC mechanism.
Hence, optimal ETC designs are only available for a limited number of settings which we explore in this paragraph.

To understand the essence of the problem, let us inspect a seminal result on this matter.
The works by \cite{astrom1999comparison}, \cite{Meng2012}, and \cite[Paper~II]{Henningsson2012} consider respectively 1-, 2-, and $n$-dimensional single-integrator systems of the form
\begin{equation}\label{eq:comparison:sys_dyn}
	\diff x(t) = u(t) \diff t + \diff v(t),
\end{equation}
where $x(t)\in\mathbb{R}^n$ refers to the system state, $u(t)\in\mathbb{R}^n$ to the control input, and $v(t)$ to an $n$-dimensional standard Wiener process.
They analyze optimality of sampling schemes with respect to the following performance measure with trade-off factor $\mu>0$
\begin{equation}\label{eq:comparison:cost_Henningsson2012}
    J = \limsup_{T\to\infty} \frac{1}{T} \E{\int_{0}^{T} x(t)^\top x(t) \diff t} + \mu \frac{1}{\E\tau},
\end{equation}
where the average sampling rate is denoted by $1/\E\tau \coloneqq \limsup_{T\to\infty}T^{-1}\sum_{k=0}^{\infty} \mathds{1}_{t_k \leq T}$ and $\mathds{1}_{(\cdot)}$ denotes the indicator function.
This yields the following theorem.

\begin{theo}{\cite[cf.\ Paper~II, Theorem~2]{Henningsson2012}}\label{thm:comparison:optimality}
    The event-triggered controller minimizing the expected average cost \eqref{eq:comparison:cost_Henningsson2012} is of an ellipsoidal absolute threshold type 
    \begin{equation*}
        t_{j+1} = \inf\left\lbrace t > t_j \mid \abs{x(t)}^2_2 \geq \sqrt{2\mu(n+2)} \right\rbrace
    \end{equation*}
    and resets the state to the origin with an impulse at each sampling instant.
    This yields the expected average cost $J = \sqrt{2n\mu}\cdot\sqrt{n/(n+2)}$.
\end{theo}

The respective optimal periodic controller utilizes analogous impulsive control inputs and yields an expected average cost $J=\sqrt{2n\mu}$.
Hence, we not only have an optimal ETC design, but also a quantification of the respective performance advantage generated by the ETC scheme.

The considered setups are restricted to specific system classes.
Hence, further research aimed to generalize the results, primarily to more general system classes and respective performance objectives.
In \cite{Molin2013}, discrete-time linear time-invariant (LTI) systems with process noise are examined with respect to a finite horizon LQR cost and a linear trade-off to the number of transmissions.
The authors arrive at certainty equivalent control laws as optimal control input.
A certainty equivalent controller employs the deterministic optimal control input while replacing the state variable by its estimate, i.e., its expected value given the past state information received at the controller.
However, the computation of an optimal ETC design requires to jointly optimize over a state estimator at the controller side and the triggering rule.
For symmetric policies, a dynamic programming approach to do so has been proposed in \cite{Molin2009}.
Closed-form solutions to the problem can however only be found for more restrictive setups.

The authors in \cite{Molin2013} study different optimal ETC design formulations:
One similar to the trade-off formulation in \eqref{eq:comparison:cost_Henningsson2012}.
Another one considering only control performance as objective, corresponding to the first term in \eqref{eq:comparison:cost_Henningsson2012}, and formulating sampling rate requirements as a constraint.
Both formulations are studied in various papers for other setups as well, e.g., in \cite{Goldenshluger2017,Soleymani2023}.

In \cite{Ramesh2013}, a similar but more general setup is studied.
The authors highlight that the control signal generally has a dual effect (of second order).
This means that it can affect not only the system state but also the future state uncertainty by purposely satisfying the triggering rule and hence closing the loop ``early''.
Hence, in principle, the controller can also steer the system such that the triggering rule is satisfied and new closed loop information is obtained ``early''.
This couples the controller, estimator, and triggering rule design, typically invalidating optimality of a certainty equivalent controller and, thus, making the design problem significantly more difficult.
The authors show that optimality of a certainty equivalent controller holds if and only if the triggering rule does not depend on the past control inputs and is hence purely innovations-based, i.e., influenced by initial condition and noise signal only.
That can however be a restrictive assumption.
Furthermore, the authors suggest a dual predictor architecture to tackle the coupled design case.

Returning to variance performance measures as in \eqref{eq:comparison:cost_Henningsson2012}, the output feedback case for continuous-time LTI systems is examined in \cite{Goldenshluger2017}.
The authors show that the controller, estimator, and triggering rule design can indeed be decoupled in this setup.
However, solving the resulting optimal stopping problem (analytically) is often difficult, and hence so is finding the optimal triggering rule.
In \cite{Andren2017}, a numerical approach is proposed to find solutions to this problem in the general case.

Considering the discrete-time linear system case with Gaussian noise and a finite horizon LQR cost as control performance measure, it is shown in \cite{Soleymani2023} that a triggering rule that uses the concept of \emph{Value of Information} is optimal.
According to the Value of Information, the system is sampled when the benefit of closing the loop outweighs its cost.
The infinite horizon case is studied in \cite{Wang2024}.
The idea also appeared in the self-triggered estimation context in \cite{Soleymani2016}.

Optimal design analyses can also be performed for estimation problems, e.g., as in \cite{Xu2004,Lipsa2011,Soleymani2016}.
Moreover, there is a limited number of publications on optimal STC design, e.g., \citep{Velasco2015,Soleymani2016}.
Furthermore, instead of considering deterministic triggering rules, stochastic sampling schemes allow for tractable closed-form analysis in some cases.
For example, in \cite{Demirel2019}, discrete-time LTI systems with Gaussian noise and output feedback are considered with a stochastic triggering rule.
Closed-form expressions of the average sampling rate and the infinite horizon average LQR cost are obtained for the optimal certainty-equivalent controller in this setting.
The authors however highlight that stochasticity of the triggering rule comes with a performance sacrifice compared to deterministic triggering rules. 

Nonetheless, it remains challenging to find and design optimal event-based control schemes.
Many traditionally introductory setups in the context of optimal control, such as LQR settings, already become challenging to analyze in the context of the performance sampling rate trade-off.
Reasons for that are, for example, the discussed dual effects and the hybrid nature of the closed loop system when introducing state-dependent triggering rules.
Hence, let us next explore a related research line aiming for a more pragmatic approach in this regard.

\subsection{Consistent design of event-based control}

Since we have seen that designing optimal event-based control schemes is very challenging in many cases, an alternative is to look for schemes that provably yield the expected trade-off advantage over periodic control.
In particular, the goal is to design event-based control schemes that result in a performance advantage compared to all periodic controllers in that setup under equal requirements on the sampling rate.
More formally, the notion of \emph{consistency} in the context of ETC has been introduced by \cite{Antunes2016}:
An event-triggering policy is (strictly) consistent if the resulting control performance is (strictly) better than that of the best periodic controller under the same average sampling rate.
Note that there is a variant of this property that additionally requires that no sampling instants are generated if no noise is present.

As this concept directly expresses a key advantage event-based control aims to achieve, it has also been considered in other works before \cite{Antunes2016}, however, in a less explicit way.
For example in \cite{astrom1999comparison}, which we have already commented on in the previous subsection, a performance advantage of $1/3$ is established for the absolute threshold triggering rule with impulsive control inputs over an impulsive (and in this setup optimal) periodic controller.
To understand the difference to the previous considerations, let us revisit the setup from the previous section \cite[Paper~II]{Henningsson2012} and formulate the respective framework:
Given \eqref{eq:comparison:sys_dyn}, compare the best periodic controller and an event-based control scheme with respect to the performance measure
\begin{equation*}
    \tilde{J} = \limsup_{T\to\infty} \frac{1}{T} \E{\int_{0}^{T} x(t)^\top x(t) \diff t}
\end{equation*}
while enforcing the constraint that the expected inter-event time is equivalent, i.e., $\tau_\mathrm{TTC} = \E{\tau_\mathrm{ETC}}$.
Then, we arrive at the following result.

\begin{theo}\label{thm:comparison:consistency}
    The ETC scheme with the absolute threshold triggering rule
    \begin{equation*}
        t_{j+1} = \inf\left\lbrace t > t_j \mid \abs{x(t)}^2_2 \geq \rho \right\rbrace,
    \end{equation*}
    and a control input that resets the state to the origin with an impulse at each sampling instant yields a performance advantage of
    \begin{equation*}
        \frac{\tilde{J}_\mathrm{ETC}}{\tilde{J}_\mathrm{TTC}} = \frac{n}{n+2},
    \end{equation*}
    for all choices of $\rho>0$ and with $\tau_\mathrm{TTC}=\E{\tau_\mathrm{ETC}}$ for the comparison.
\end{theo}
\begin{proof}
    This can be deduced from the results in \cite[Paper~II]{Henningsson2012} or directly be seen from \cite[Prop.~2]{Goldenshluger2017} for $\Sigma_v,\Sigma_u \to 0$.
\end{proof}

Note that, by inserting $n=1$, we obtain the result from \cite{astrom1999comparison}.
Furthermore, this result matches our expectation that the optimal ETC scheme described in the previous subsection generates a performance advantage when compared to the best periodic controller.
However, we have now obtained a framework that does not require us to solve the optimal design problem for ETC, but instead allows us to evaluate the benefits of suboptimal event-based control designs as well.
In \cite{Goldenshluger2017}, this is explicitly pointed out, stating that a pragmatic approach to circumventing the optimal stopping problem is to find schemes that outperform the best periodic controller for the same expected inter-event time.

A simulation-based consistency examination for a self-triggered discounted linear quadratic control problem can be found in \cite{gommans2014self}.
Consistent ETC design schemes for various setups have been considered in \cite{Antunes2018,Balaghiinaloo2021,Balaghiinaloo2022}.
These works have considered discrete- and continuous-time setups, mostly for linear (stochastic) systems, and consistency with respect to different performance measures yielding (in parts strict) LQ-, or $\ell_2$-consistency.
A consistent absolute threshold triggering rule for a periodic ETC scheme and linear stochastic system is presented in \cite{Khashooei2018}.
In \cite{Brunner2018}, the authors propose an infinite horizon LQ-consistent ETC design for a stochastic triggering rule for discrete-time LTI systems under Gaussian noise and state feedback, preceding the optimal controller design for the analogous output feedback setup in \cite{Demirel2019}.
Strict LQ-consistency of an event-based control scheme for an LQG setup is examined in \cite{Yu2022}.
Moreover, in the $H_\infty$ context, consistency is shown in \cite{Mi2022}.

These works demonstrate the potential of event-based control to yield a performance advantage over periodic control and, hence, for example, its ability to reduce the average sampling rate under the same performance requirements.
Various interpretations of consistency or its absence are discussed in \cite{Meister2025}, including a discussion on how to compare sampling schemes.
However, the advantage of event-based control over periodic control cannot be taken for granted and should ideally be shown for proposed event-based control designs.
For example, in \cite{Meister2023}, the absolute threshold triggering rule from Theorems~\ref{thm:comparison:optimality} and \ref{thm:comparison:consistency} is shown to become inconsistent when the vector 2-norm in the triggering rule is replaced by the vector maximum norm.
Moreover, in \cite{Antunes2023}, the authors show that, for a discrete-time, linear time-invariant system setup, there always exists a disturbance such that no \emph{strictly} $\ell_2$-consistent ETC scheme can be designed.
As pointed out in \cite{Nowzari2019}, there is currently a lack of results analyzing the potential advantage of event-based control over TTC.
A starting point for this is \cite{Meister2024}, where the (decentralized) absolute threshold triggering rule is studied in a consensus problem that is otherwise analogous to the setups in \cite{astrom1999comparison} and \cite[Paper~II]{Henningsson2012}.
Surprisingly, the consistency property of the ETC scheme we got to know in Theorem~\ref{thm:comparison:consistency} vanishes in this analogous setup.
Consequently, performant centralized designs of ETC schemes do not naturally perform well in decentralized cases, and it remains to be examined what performant designs for distributed settings are.

\subsection{Perspectives for further research}

Up until today, the event-based community has managed to find design schemes for a limited number of settings that 
(a) are optimal with respect to the performance sampling rate trade-off, or
(b) provably outperform the best periodic controller in the consistency sense.
However, many other existing schemes have not been analyzed rigorously regarding their performance potential, also cf.\ \citep{Goldenshluger2017,Nowzari2019}.
Thus, the need for analyzing performance properties of existing and newly proposed event-based control schemes is pressing.

Hence, a research direction with high impact potential is to bring together the performance related techniques elaborated in this section with the event-based control frameworks examined in Section~\ref{sec_framework}.
This would allow obtaining design frameworks that broadly characterize the performance properties of resulting event-based control schemes and, thereby, more widely enable performance advantages of event-based control by design.

Another promising research direction is to ease the analysis of specific event-based control designs with respect to their performance properties.
Since optimal designs are typically hard to find, studying the consistency properties of existing designs in a broader way has a lot of potential.
We have seen that control performance requirements and resulting sampling behavior are strongly intertwined for event-based control schemes in many cases.
Consequently, one can aim to find new theoretical approaches for this analysis, but may very well also focus on numerical and algorithmic consistency evaluations of event-based control schemes.
For this goal, one could leverage the (numerical) sampling behavior analysis techniques from Chapter~\ref{sec_sampling} to arrive at methodologies that allow for (numerical) performance evaluations for existing ETC schemes.
This could ultimately provide a framework to analyze newly designed triggering rules (numerically) and to relate existing and new event-based control schemes in terms of their performance properties.
Performance comparison perspectives for sampling schemes and their relationship are discussed in \cite{Meister2025} and can serve as a starting point for such (numerical) tools.

Lastly, the concept of ETC performance examination and how to compare it to TTC should be transferred to different application regimes of ETC.
As an example, the advantage of decentralized ETC over periodic schemes in distributed setups is mostly unexamined, as pointed out above and in \cite{Nowzari2019}.
A potential research direction is to study ETC designs that take the availability of information explicitly into account.
In \cite{Meister2024a}, it is demonstrated that this can have decisive impact on ETC performance properties in distributed setups.
Other application domains may also require analyses with respect to other performance criteria or different sampling intensity measures, e.g., the minimum inter-event time or the data rate.
Details on the data rate as sampling cost and its motivation can be found in the next section where we move from control performance to closed-loop stability as control goal.

%% file: information.tex
\section{Fundamental limitations for rate-limited channels}
\label{sec_info}
Event-based control is often considered in setups where the sensors and actuators of the plant are connected by a rate-limited digital communication channel.
In such setups, not only the number of sampling instants but also the amount of information (e.g. in terms of bits) that is transmitted at sampling instants needs to be taken into account.
A question that arises naturally in this context is which minimum data rate is required for systems with ETC and how this minimum rate relates to the one required for TTC.
In this section, we hence discuss how the data-rate requirements for ETC can be related to the requirements for TTC.
Fur this purpose, we pair the data rate as a resource utilization measure with stabilization (or convergence rate) as a control goal.

We restrict ourselves to the linear setup with $f(x,\hat{u}) = Ax+B\hat{u}$ in \eqref{eq_plant}, with $\abs{x(0)}_2<L$ for some known constant $L>0$.
Different from the standard setup presented in Section~\ref{sec_sub_setup}, the coding between sensors and actuators needs to be considered for the controller.
We thus assume that the controller is attached to the actuators of the plant and has the form $\hat{u} = \kappa(\hat{x})$.
Here, $\hat{x}$ is a reconstruction of the plant state that depends on information that has been received over the channel.
How $\hat{x}$ is reconstructed depends on the specific coding scheme that is used.

In line with \cite{Khojasteh2016}, we distinguish between information access rate and information transmission rate for the presented data rate results.
The \emph{information access rate} (IAR) determines the data rate at which the controller receives information from the plant.
We denote by $b_c(t)$ the amount of information in bits received from the plant (and hence accessed) by the controller up to time $t$.
The information access rate can be defined as
\begin{equation*}
	R_c = \limsup_{t\rightarrow\infty} \frac{b_c(t)}{t}.
\end{equation*}
In turn, the \emph{information transmission rate} (ITR) refers to the average data payload per time interval sent by the sensor over the channel.
Let $b_s(t)$ denote the number of bits in the data payload transmitted over the channel up to time $t$.
Then, the ITR can be defined as
\begin{equation*}
	R_s = \limsup_{t\rightarrow\infty} \frac{b_s(t)}{t}.
\end{equation*}
Note the subtle difference between the two information rates:
While the ITR only considers the actual number of bits transmitted via the channel, the IAR takes into account the information from the plant received through data payload as well as implicitly, e.g., from the timing of received data packets.
The IAR is independent of the form in which the information is received, i.e., as data payload or encoded in the timing of a data packet, whereas the ITR only considers the data payload.
The required IAR to achieve certain control goals, like stabilization, is hence independent of the deployed sampling scheme or the presence of delays.
On the contrary, this is not the case for the ITR which depends on such setup aspects.
Consequently, we first present a data-rate theorem which governs the required IAR for stabilization and holds for TTC and ETC alike.
Then, we present ITR results in two subsections on TTC and ETC and relate them to each other.

\subsection{Information access rate requirements}

The IAR is governed by the following data-rate theorem.
\begin{theo}\cite[Theorems~1 and 2]{hespanha2002towards}.
	\label{theo_iar}
	The IAR required to stabilize system~\eqref{eq_plant} for $f(x,\hat{u}) = Ax+B\hat{u}$ over a digital channel satisfies
	\begin{equation}
		\label{eq_info_rmin}
		R_c \geq r_{\mathrm{IAR}}(A) \coloneqq \frac{1}{\ln(2)} \sum_{i:\Re{\lambda_i(A)} > 0} \lambda_i(A),
	\end{equation}
	where $\lambda_i(A)$ denotes the $i$-th eigenvalue of $A$.
	If $A$ is diagonalizable, then this stabilization is possible with an IAR equal to $r_{\mathrm{IAR}}(A)$.
\end{theo}
Theorem~\ref{theo_iar} thus implies that no stabilization is possible with an IAR $R_c < r_{\mathrm{IAR}}(A)$, i.e. the rate $r_{\mathrm{IAR}}(A)$ is necessary.
Moreover, a specific optimal coding scheme to achieve stabilization at rate $r_{\mathrm{IAR}}(A)$ is presented in \cite{hespanha2002towards}.
Thus, the IAR $r_{\mathrm{IAR}}(A)$ is also sufficient.
The bound $r_{\mathrm{IAR}}(A)$ is also referred to as inherent entropy rate of the system.
Note once more that this result is independent of the presence of communication delays or the deployed sampling scheme.

If we additionally aim for exponential convergence of the state to the origin, an extension of this data-rate theorem (with a necessary condition) can be found in \cite{khojasteh2020value}.
Further data-rate theorems for other setups, e.g., nonlinear systems, systems with uncertain parameters, stochastic noise channels, or time-varying channels, can be found in the literature.
Introductory overviews can, for example, be found in \cite{Nair2007,Franceschetti2014}.

\subsection{Information transmission rate requirements}

Let us now inspect ITR results for TTC and ETC under unknown communication delays, derived in \cite{Khojasteh2017,khojasteh2020value}.
To keep the notation simple, we restrict ourselves to the scalar case with $n_x = n_u = 1$ and $A > 0$ for the rest of this section.
Extension to higher dimensions can be found in \cite{Khojasteh2017,khojasteh2020value}.

As we consider unknown delays, we have to distinguish between time instants when packets are transmitted and time instants when packets are received.
We denote by $(t_j^s)_{j\in\N_0}$ the sequence of transmission time instants and by $(t_j^c)_{j\in\N_0}$ the sequence of time instants when packets are received on the controller/sensor side.
A packet sent at time $t_j^s$ is received at time $t_j^c$ and thus the (unknown) communication delay can be computed as $\Delta_j = t_j^c-t_j^s$.
We assume that the communication delays are bounded by some known constant $\boldsymbol{\Delta}>0$, i.e., $\Delta_j \leq \boldsymbol{\Delta}$ for all $j\in\N_0$.

For the considered TTC and ETC schemes, the estimate $\hat{x}$ is designed as
\begin{subequations}
	\begin{align}
		\dot{\hat{x}}(t) ={}  & A\hat{x}(t) + B \hat{u}, \quad t\in \left[t_j^c,t_{j+1}^c\right)                  \\
		\hat{x}(t_j^{c+}) ={} & \hat{x}(t_j^c) + \mathfrak{g}(x(t_j^s),\hat{x}(t_j^s),t_j^s). \label{eq_info_est}
	\end{align}
\end{subequations}
Recall that, in this section, the input  that is applied to the plant is determined as $\hat{u}  = \kappa(\hat{x})$.
Here the function $\mathfrak{g}:\R^3 \rightarrow \R$ describes the update of $\hat{x}$ based on information received through the channel.
Such information can include quantized state information as well as quantized information about the timing of sampling instants.
It thus depends on the state $x(t_j^s)$, the estimate $\hat{x}(t_j^s)$, the triggering instant $t_j^s$, and the used coding scheme.
It is assumed in \cite{Khojasteh2017,khojasteh2020value} that the sensor can reconstruct the estimate $\hat{x}$, which can for example be inferred from the input applied to the plant.
The latter can be reconstructed if measurements of $\dot{x}$ are available and $\kappa$ is invertible.
Using $\hat{x}$, the state estimation error can then be computed as
\begin{equation*}
	z(t) = x(t) - \hat{x}(t)
\end{equation*}
with the initial condition $z(0) = x(0) - \hat{x}(0)$.

\subsubsection{ITR requirements for TTC}

Considering a periodic TTC scheme with sampling instants $t_j^s = j\tau$ and constant transmission period $\tau>0$, the following theorem can be formulated.
\begin{theo}\cite[Theorem~2]{Khojasteh2017}.
	\label{theo_itr_periodic}
	Consider system~\eqref{eq_plant} for $f(x,\hat{u}) = Ax+B\hat{u}$ being controlled over a digital channel with delays bounded by some known constant $\boldsymbol{\Delta}>0$ and packet order-preserving transmission and decoding.
	Then, there exists a delay realization $(\Delta_j)_{j\in\mathbb{N}_0}$ such that an ITR
	\begin{equation*}
		R_s \geq \begin{cases}
			\frac{A}{\ln(2)}, \quad                                        & \text{if $\boldsymbol{\Delta} < \tau$}, \\
			\frac{A}{\ln(2)} \cdot \frac{\boldsymbol{\Delta}}{\tau}, \quad & \text{otherwise},
		\end{cases}
	\end{equation*}
	is necessary for stabilization when considering the periodic TTC scheme.
\end{theo}
Note that this result only provides a necessary condition for stabilization under unknown delays.
For sufficiently small delay bounds, the necessary ITR is equivalent to the IAR formulated in Theorem~\ref{theo_iar}.
When delays larger than the transmission period $\tau$ can occur, the required ITR for stabilization increases linearly.
From a qualitative perspective, sufficiently large delays reduce the information about the plant contained in the data payload transmitted via the channel since this information is outdated.
Hence, in order to achieve the required IAR for stabilization, an increased ITR is required.
This again highlights the necessary distinction between IAR and ITR.
Moreover, without delays, the required IAR and ITR are equivalent.
This is to be expected as, for TTC schemes, the timing of data packets does not bear information.

Let us now contrast this result to an ITR result for ETC.

\subsubsection{ITR requirements for ETC}

An important motivation for ETC is that it is more efficient regarding the use of communication resources compared to TTC.
To actually confirm this claim, an important research question is whether systems can be stabilized using ETC with a lower ITR than required for the TTC scheme in Theorem~\ref{theo_itr_periodic}.
It turns out that, in a setup where no communication delays occur, it is even possible to stabilize linear systems with ETC with an ITR of zero.
In particular, an ETC scheme based on deadbeat control that requires only the transmission of $2n_x+2$ bits is presented in \cite{kofman2006level}.
For this approach, a one dimensional output to the plant is considered and one bit is transmitted whenever the output of the plant changes by a constant value, that is known both to the coder and the decoder.
After $2n_x+2$ transmissions of that kind, the state of the plant can be reconstructed given that the plant is observable from the considered output.
Then a deadbeat controller can be used to steer the state to the origin in finite time.
As noted in \cite{kofman2006level}, at first glance, this ETC scheme seems to be contradictory to the data-rate theorem in Theorem~\ref{theo_iar}.
However, with the distinction between IAR and ITR, it is clear that this is not the case.
In fact, stabilizing the system with a finite number of received bits is possible, since the timing of the trigger events bears additional information.
As soon as unknown\footnote{In case the delays are known, the timing information can still be used to reconstruct the state.} delays occur, which are in practice unavoidable in channels with rate limitations, the ETC scheme from \cite{kofman2006level} is no longer applicable since the exact timing information is lost.

However, the timing of the triggering events does still bear information when unknown delays occur.
This is analyzed thoroughly for a specific ETC scheme in \cite{khojasteh2020value}.
Therein, necessary and sufficient conditions for the ITR required for stabilization and exponential convergence (of the state estimation error $z(t)$) with the considered ETC scheme are derived.
In order to illustrate effects that can occur for ETC over a rate-limited channel, we sketch these results here.

The transmission instants are determined using an ETC triggering rule with decreasing threshold.
The threshold function is given by
\begin{equation*}
	v(t) = v_0 \e^{-\psi t}
\end{equation*}
for $v_0 > 0$ and $\psi > 0$.
A transmission is triggered as soon as $z(t)$ reaches the value of $v(t)$.
Formally, the triggering rule can be stated as
\begin{equation}
	\label{eq_trigger_info}
	t_{j+1}^s = \inf\left\lbrace t > t_j^c\mid \abs{z(t)} \geq v(t) \right\rbrace.
\end{equation}
Note that the triggering rule ensures that a triggering event only occurs after the previous transmission has been received.
This requires suitable sensor measurements to infer whether the previous state has been received.
Considering \eqref{eq_info_est},
note that $\mathfrak{g}(x(t_j^s),\hat{x}(t_j^s),t_j^s)$ should yield a quantized estimate of the value of $z(t_j^c)$ based on the information received through the channel.
Then, $z(t_j^{c+})$ becomes small.
Moreover, the quantization in $\mathfrak{g}(x(t_j^s),\hat{x}(t_j^s),t_j^s)$ must allow the estimation error $z$ to drop below the triggering threshold when a new packet is received over the channel.
This is needed to guarantee a positive inter-event time.
In particular, we assume that the quantization is such that
\begin{equation}
	\label{eq_info_quant}
	\begin{split}
		\abs{z(t_j^{c+})} ={} & \abs{z(t_j^c) - \mathfrak{g}(x(t_j^s),\hat{x}(t_j^s),t_j^s)}              \\
		\leq{}                & \varrho_0 \e^{-\psi \boldsymbol{\Delta}} v(t_j^s) \leq \varrho_0 v(t_j^c)
	\end{split}
\end{equation}
holds for some $\varrho_0\in\left(0,1\right)$.
This can be achieved by quantization with sufficiently large, finite precision that we denote subsequently by $\nu > 1$.
The quantizer precision $\nu$ describes essentially the inverse factor by which the uncertain region that contains the state shrinks due to a transmission, see \cite{khojasteh2020value}.
For any coding scheme that satisfies \eqref{eq_info_quant}, the following statements on the data rate required for stabilization can be made.

\begin{theo}\cite[Theorem~2]{khojasteh2020value}.
	\label{theo_itr_etc_nec}
	If \eqref{eq_info_quant} holds with a quantizer of precision $\nu$, then there exists a delay realization $(\Delta_j)_{j\in\N_0}$ and an initial condition such that the ITR satisfies
	\begin{equation*}
		R_s \geq \frac{A + \psi}{\ln(\nu) + \ln(2+\frac{\e^{\psi \boldsymbol{\Delta}}}{\varrho_0})} \max \left\lbrace 0, \log(\frac{\e^{A \boldsymbol{\Delta}} -1}{\varrho_0 \e^{-\psi \boldsymbol{\Delta}}})\right\rbrace
	\end{equation*}
	in the considered setup for any coding scheme when using the proposed ETC scheme.
\end{theo}
Theorem~\ref{theo_itr_etc_nec} thus provides a lower bound on the ITR required for the proposed ETC scheme, hence a necessary condition for stabilization and exponential convergence of the estimation error.
The proof of Theorem~\ref{theo_itr_etc_nec} works by deriving a lower bound on the number of bits that the sensor needs to transmit at each triggering event as well as a lower bound on the sampling rate.

In \cite{khojasteh2020value}, the authors also derive a sufficient condition for the ITR under ETC.
\begin{theo}\cite[Theorem~3]{khojasteh2020value}.
	\label{theo_itr_etc_suff}
	If the state estimation error satisfies $\abs{z(0)}<v_0$, then for any ITR
	\begin{align*}
		R_s \geq{} & \frac{A + \psi}{-\ln(\varrho_0 \e^{-\psi \boldsymbol{\Delta}})}                                                                            \\
		           & \cdot \max \left\lbrace 0, 1 + \log(\frac{bA (\boldsymbol{\Delta}+\psi)}{\ln(1+\varrho_0\e^{-(A+\psi)\boldsymbol{\Delta}})})\right\rbrace,
	\end{align*}
	where $b>1$, there exists a quantization strategy that achieves \eqref{eq_info_quant} in the considered setup when using the proposed ETC scheme.
\end{theo}

Let us investigate the bounds on $R_s$ in Theorems~\ref{theo_itr_etc_nec} and \ref{theo_itr_etc_suff} for $\psi \rightarrow 0$, i.e., if we only consider stabilization under an absolute threshold triggering rule.
We can then relate them to the ITR result for TTC and the IAR result.
For Theorem~\ref{theo_itr_etc_nec}, let us additionally consider $\varrho_0 \ll 1/\max\{2,\nu\}$ for illustrative purposes.
The necessary condition on the ITR in the ETC case from Theorem~\ref{theo_itr_etc_nec} simplifies to
\begin{equation*}
	R_s \geq \frac{A}{\ln(2)} \max\left\lbrace 0, 1 + \frac{\log(\e^{A\boldsymbol{\Delta}}-1)}{-\log(\varrho_0)} \right\rbrace.
\end{equation*}
We can conclude in this case that the necessary ITR for the considered ETC scheme matches $r_\mathrm{IAR}(A)$ from Theorem~\ref{theo_iar} for $\boldsymbol{\Delta} = \ln(2)/A$, namely $R_s = A/\ln(2)$.
This is the point at which the information loss due to delays is exactly compensated by the additional timing information from ETC.
The result is therefore equivalent to the one without delays and timing information and also matches the ITR bound for TTC in Theorem~\ref{theo_itr_periodic}.
With the help of the sufficient ITR condition in Theorem~\ref{theo_itr_etc_suff}, we can in principle investigate when the ETC scheme offers a data rate advantage over the TTC scheme by comparing to Theorem~\ref{theo_itr_periodic}.
Omitting a quantitative analysis, we can assess that the proposed ETC scheme offers a data-rate advantage over the TTC scheme for sufficiently small delay bounds.
However, there is no conclusive picture for larger delay bounds and further investigations, potentially including other sufficient conditions for the ETC ITR are needed.
The superiority of ETC schemes over periodic TTC schemes in terms of the data rate can therefore not be concluded in general.
It is hence worth investigating this relationship further.

It is important to note that all statements in this section are limited to the specific ETC scheme and setup considered in \cite{khojasteh2020value}.
Nonetheless, there are other results in the literature that align with the qualitative findings in this section.
It is reported in \cite{linsenmayer2017delay} that uncertainty in the timing information generates increased ITR requirements for containability of scalar systems with ETC.
Moreover, \cite{ling2017bit} studies sufficient ITR conditions for ISS of an ETC scheme similar to the one in this section.
The authors show that the ETC scheme can stabilize a scalar system with a smaller ITR than TTC in certain situations, again highlighting the importance of timing information in ETC.
Further results where lower bounds on data rates required for ETC are studied can be found, e.g., in  \cite{li2012stabilizing,tallapragada2016event,linsenmayer2018containability}.

\subsection{Perspectives for further research}

In this section, we have seen that for certain setups the answer to whether ETC or TTC allows a lower data rate is not trivial to answer, for example due to the ITR requirements depending on the maximum delay that can occur.
There are setups in which certain ETC schemes are provably preferable for data-rate efficient control, particularly under little timing uncertainty, i.e., small unknown delays.
But universal superiority of ETC in practical setups, including for example significant communication delays, cannot be easily concluded.
From a data-rate perspective, it is therefore necessary to analyze exactly which sampling strategy is better for each setup.
However, the existing analysis tools are often restrictive and cover only simple (often scalar) setups and specific triggering rules.
An evident goal for future research is therefore to develop methods for more general setups and triggering rules.
Specifically, developing results for linear systems for other standard triggering rules and non-scalar setups are meaningful next steps.
For nonlinear systems, developing qualitative results that relate the performance of ETC to TTC may be more difficult, since the required bit rate may vary depending on the region of the state space where a system is operating. 
Such effects were, e.g., reported \cite{li2012stabilizing}.
ETC may thus offer more flexibility to adjust the data rates for nonlinear systems.

The use of timing information opens up new perspectives for ETC.
As seen in this section, using such information is clearly beneficial for designing data-rate efficient ETC triggering mechanisms.
However, it still needs to be investigated how the channel can be used in an optimal way.
An overarching goal can thus be to develop data-optimal ETC strategies with optimal channel usage.

%% file: further_trends.tex
\section{Further research trends}
\label{sec_current}
In this section, we give an overview of further trends in the current research on event-based control and sketch possible future research directions. 
Note that we cannot provide an exhaustive list of references for each trend in this section but rather point out ideas with reference examples as starting points for further investigation.

The development of new classes of ETC triggering rules and new analysis techniques for existing ones remains a relevant topic. 
In \cite{kurtoglu2023energy}, an ETC triggering rule for linear systems is derived based on an energy function that captures the difference to a set point.
The triggering rule is norm free in the sense that it compares functions that can take negative values and involves also the time derivative of an energy function. 
A triggering rule based on a performance barrier function is proposed in \cite{ong2024performance}. 
A looped function approach is used to derive static and dynamic triggering rules in \cite{xu2024aperiodic}.
Positive systems are leveraged in \cite{alyahia2024dynamic} in the design of triggering rules. 
Conceptually novel ETC triggering rules not only have the potential to lead to better closed-loop behavior but also to improve the understanding of ETC and its properties.
This is a reason why research efforts should continue to focus on developing new triggering rules, as long as such triggering rules offer conceptual novelties that set them apart from the existing literature. 
In addition, the development of a benchmark that allows different ETC concepts to be compared in a meaningful way is a relevant topic for future research.
Such comparisons will involve the trade-off between achieving a control objective, e.g., stability in the closed loop or a certain control performance level, and the required sampling cost, e.g., in terms of the average sampling- data rate, see also \cite{heemels2012introduction}.
Instead of numerical or analytical approaches to compare ETC schemes, a benchmark can help to obtain reliable simulation results as another angle of attack.
Inspired by consistency discussed in Section~\ref{sec_comparison}, comparing to (optimal) periodic control schemes in these benchmark scenarios can be a good reference for different ETC schemes as well.

A number of papers combine other techniques that are relevant in current research with event-based control. 
This either aims at improving the performance of event-based control techniques or to apply event-based control techniques to various problems where it can help to develop novel solution approaches.
In \cite{rajan2024event}, event-triggered parameterized control is studied. 
Therein, instead of holding the input constant between sampling instants, the input is parameterized using a combination of a set of basis functions.
This yields an alternative to model-based ETC with reduced computational complexity. 
Another recent topic is the safety of systems with ETC, that can be guaranteed using barrier functions \citep{taylor2021safety,xiao2021event} and safety filters \citep{ong2024hierarchical}.

In setups with shared resources such as NCS, resource usage cannot be performed opportunistically but needs to be coordinated.
Opportunistic sampling, often shown by ETC schemes, can lead to inefficient resource usage and even control performance degradation, see, e.g., \cite{Blind2013}.
\cite{Blind2013} demonstrate the impact of various medium access control protocols on control performance for an absolute threshold ETC scheme and a periodic control scheme.
An alternative to centralized medium access protocols are decentralized traffic specifications for resource coordination.
In \cite{antunes2014rollout}, rollout ETC is deployed with a window-based traffic specification.
In \cite{wildhagen2019predictive,wildhagen2023uncertainties}, rollout ETC is combined with a token bucket traffic specification. 
Token bucket models are frequently used for traffic specifications in computer networks.
Compared to window-based traffic specifications, they come with higher complexity but more scheduling flexibility.
The resulting ETC approach in \cite{wildhagen2019predictive,wildhagen2023uncertainties} allows efficient scheduling for networks that include a token bucket traffic specification.
Moreover, STC offers a natural way of alleviating greedy sampling while sacrificing some reactivity.
We contrast ETC and STC in this regard at the end of this section.
Practical implementation of event-based control in the NCS context has been considered in the development of network stacks and architectures able to cope with the aperiodic nature of different sampling strategies, e.g., \cite{Araujo2014,Trobinger2021,Trobinger2024,Laidig2024}.

Another trend connected to NCS is the development of cybersecure and privacy-preserving ETC and STC schemes.
As motivated in \cite{Shoukry2016}, resilience to cyberattacks becomes a key factor in industry and, thus, the authors propose an event-triggered observer resilient to sparse sensor attacks.
Resilience of ETC schemes to denial of service attacks is considered in various settings, e.g., \cite{ShishehForoush2012,dolk2017denial,Zhao2022}.
Moreover, watermarking strategies allow verifying the authenticity of communicated messages in ETC \citep{Barboni2022} ans STC schemes \citep{Wolleswinkel2024}.
While the typically sparse communication pattern of ETC and STC can have advantages with respect to cyberattacks, there is additional information contained in the sampling pattern, as also alluded to in Section~\ref{sec_info}.
Consequently, the sampling pattern can reveal information about the closed loop which may endanger privacy of closed loop properties and states, see, e.g., \cite{Sharma2021}.
Hence, the development of privacy-preserving sampling schemes attracts research interest in this domain as well, e.g., \cite{Rikos2023}.

There is also recent literature on combining event-based control 
and data-driven control. 
In \cite{wang2023model,dePersis2024event}, approaches to design ETC for unknown linear systems based on noisy state measurements are presented. 
A data-based analysis tool for the sampling performance of systems with unknown dynamics is proposed in \cite{peruffo2024sampling}.
Moreover, ETC also has the potential to complement data-driven control approaches.
In \cite{iannelli2024hybrid}, it is shown how ETC can be used in data-driven control for time-varying systems to identify time instants when new data needs to be recorded. 
This enables efficient and sustainable data collection. 

Learning ETC triggering rules based on reinforcement learning is studied in \cite{baumann2018deep,funk2021learning}.
In \cite{ornia2022event}, ETC is leveraged to reduce the amount of communication in distributed Q-learning. 
The concept of event-triggered learning is introduced in \cite{solowjow2020event}. 
In event-triggered learning, a triggering rule is used to decide when a new model needs to be learned for making predictions of time-varying systems.
Overall, it can be summarized that there are now various examples where event-based control complements other technologies. 
In this way, the advantages of  event-based control can be utilized for a variety of applications, rendering them for example more data-, energy-, or computationally efficient.

Regarding the literature for STC, it can be concluded that there exist various approaches to derive triggering rules. 
However, it is generally not clear a priori which approach leads to the best closed-loop behavior. 
This should therefore be analyzed in the future, particularly when new approaches are proposed. 
Similar problems arise in this context as for the analysis of ETC, so that a good benchmark might also be beneficial to that end. 

Furthermore, it should be examined further, when STC and when ETC is actually the better choice. 
In general, STC often has the advantage that it is better suited for scheduling, as sampling instants are already known in advance.
On the other hand, ETC has the advantage that it is better able to react to disturbances, that can be taken into account immediately for ETC whilst these  can only be taken into account at the next sampling instant for STC. 
To make these effects clearer, it would be interesting to further investigate the application of STC in real world setups. 
This was done for the optimization of CPU usage in \cite{samii2010dynamic} and for a simple networked control setup in \cite{camacho2010self}. 
However, also for more realistic networked control setups and for more complex system dynamics, such an analysis would be interesting, and particularly a comparison of the performance of ETC and STC might help to evaluate both.

%% file: coclusion.tex
\section{Conclusion}
\label{sec_conc}

In this paper, we have provided an introduction to and an overview of existing research on the theory of event-based control.
Moreover, we have presented current research trends in the field which come with pressing open questions and research perspectives for the field.
In particular, the following trends have been extensively discussed:
\begin{itemize}
    \item the development a unifying and general theory for event-based control to facilitate the analysis and design of triggering rules and improve the understanding of underlying mechanisms in event-based control,
    \item the creation of analysis tools for the sampling behavior of event-based control, i.e., to analyze what sampling patterns or inter-event time properties, e.g., average inter-event times, occur for closed-loop systems with event-based control,
    \item the analysis of the performance sampling rate trade-off for event-based control and, thereby, enabling the comparison of  event-based control approaches among each other and to TTC with respect to control performance and sampling efficiency, and
    \item the examination of fundamental limitations of event-based control for data-rate limited channels, i.e., when considering the data rate instead of the sampling rate as a resource.
\end{itemize}
Progressing in these areas will provide a broadly applicable design framework for event-based control schemes, resource usage quantification tools usable for system design, methods to judge the control and sampling performance of event-based control schemes making them comparable, and an understanding of data-efficient communication.
An important goal for future system theoretical research is therefore to address the open questions identified in this work in order to better understand the benefits of event-based control and how to leverage them.

Besides this important aspect, we see it as an evident next step to transfer the concepts for event-based control to other domains than classical networked control or scheduling for real-time systems. 
For example, event-based control has great potential to provide advantages in applications such as event-based vision, where pixels are updated according to triggering rules in order to gain various computational benefits \citep{gallego2022event}.
Another example is neuromorphic control, where the potential of event-based control was recognized early \citep{deweerth1991simple} and for which it has recently been demonstrated that event-based control can enable energy efficient controller designs \citep{schmetterling2024neuromorphic}.
Event-based control may also be useful in predictive maintenance, where increasing amounts of data pose challenges of data-driven modeling \citep{nunes2023challenges}. 
Many other applications can be envisioned as well. 
To leverage the full potential of event-based control, it is therefore important to find ways to make profitable use of the existing theory and to develop the theory accordingly, where gaps still exist.